\documentclass[compress,final,3p,times,12pt]{elsarticle}
\usepackage{amsfonts}
\usepackage{graphics}
\usepackage{amssymb}
\usepackage{amsmath}
\usepackage{array}
\usepackage{amsmath}
\usepackage{amsfonts}
\usepackage{amssymb}
\usepackage{algorithm}
\usepackage{algorithmic}
\usepackage{graphicx}
\usepackage{verbatim}

\newtheorem{theorem}{Theorem}
\newtheorem{lemma}{Lemma}
\newtheorem{remark}{Remark}

\newtheorem{corollary}{Corollary}
\newtheorem{example}{Example}
\newtheorem{definition}{Definition}

\newcommand{\beq}{\begin{equation}}
\newcommand{\eeq}{\end{equation}}
\newcommand{\beqnn}{\begin{equation*}}
\newcommand{\eeqnn}{\end{equation*}}
\newcommand{\beqy}{\begin{eqnarray}}
\newcommand{\eeqy}{\end{eqnarray}}
\newcommand{\beqynn}{\begin{eqnarray*}}
\newcommand{\eeqynn}{\end{eqnarray*}}
\newcommand{\bit}{\begin{itemize}}
\newcommand{\eit}{\end{itemize}}
\newcommand{\ben}{\begin{enumerate}}
\newcommand{\een}{\end{enumerate}}
\newcommand{\bex}{\begin{example}}
\newcommand{\eex}{\end{example}}

\newcommand{\balg}[1]{\begin{algorithm} \caption{#1}}
\newcommand{\ealg}{\end{algorithm}}

\newcommand{\balgc}{\begin{algorithmic}[1]}
\newcommand{\ealgc}{\end{algorithmic}}

\newcommand{\bary}{\begin{array}}
\newcommand{\eary}{\end{array}}
\newcommand{\bmx}{\begin{bmatrix}}
\newcommand{\emx}{\end{bmatrix}}
\newcommand{\bsmx}{\left[\begin{smallmatrix}}
\newcommand{\esmx}{\end{smallmatrix}\right]}
\newcommand{\bmxc}[1]{\left[\begin{array}{@{}#1@{}}}
\newcommand{\emxc}{\end{array}\right]}
\newcommand{\bcn}{\begin{center}}
\newcommand{\ecn}{\end{center}}

\newcommand{\A}{\boldsymbol{A}}
\newcommand{\B}{\boldsymbol{B}}
\newcommand{\C}{\boldsymbol{C}}

\newcommand{\E}{\boldsymbol{E}}

\newcommand{\I}{\boldsymbol{I}}

\renewcommand{\P}{\boldsymbol{P}}

\newcommand{\e}{\boldsymbol{e}}

\newcommand{\h}{\boldsymbol{h}}

\newcommand{\q}{\boldsymbol{q}}
\newcommand{\rr}{\boldsymbol{r}}

\renewcommand{\u}{\boldsymbol{u}}
\renewcommand{\v}{\boldsymbol{v}}
\newcommand{\w}{\boldsymbol{w}}
\newcommand{\x}{{\boldsymbol{x}}}
\newcommand{\y}{{\boldsymbol{y}}}

\newcommand{\0}{{\boldsymbol{0}}}

\newcommand{\1}{{\boldsymbol{1}}}

\numberwithin{equation}{section}
 \numberwithin{Lem}{section}
 \numberwithin{Defi}{section}
 \numberwithin{Theo}{section}
 \numberwithin{Rem}{section}
  \numberwithin{Coro}{section}
  \numberwithin{Fig}{section}

\include{micros}

\journal{}

\begin{document}

\begin{frontmatter}



\title{Sharp Sufficient Conditions for Stable Recovery of Block Sparse Signals  by Block Orthogonal Matching Pursuit\tnoteref{label1}} 

 \author{Jinming Wen\fnref{addr1}}
\ead{jinming.wen@mail.mcgill.ca}
\author{Zhengchun Zhou\fnref{addr2}}
\ead{zzc@home.swjtu.edu.cn}
 \author{Zilong Liu\fnref{addr3}}
\ead{zilong.liu@surrey.ac.uk}
 \author{Ming-Jun Lai\corref{cor1}\fnref{addr4}}
\ead{mjlai@uga.edu}\cortext[cor1]{Corresponding author}
 \author{Xiaohu Tang\fnref{addr5}}
\ead{xhutang@swjtu.edu.cn}

\address[addr1]{College of Information Science and Technology, Jinan
University, Guangzhou, 510632, China.}
\address[addr2]{School of Mathematics, Southwest Jiaotong University,
Chengdu 610031, China}
\address[addr3]{ Institute for Communication Systems (ICS), 5G Innovation Centre (5GIC), University of Surrey, UK}
\address[addr4]{Department of Mathematics, University of Georgia, Athens, GA 30602, U.S.A.}
\address[addr5]{Information Security and National Computing Grid
Laboratory, Southwest Jiaotong University, Chengdu 610031, China}

\begin{abstract}
In this paper, we use the block orthogonal matching pursuit (BOMP) algorithm to recover block sparse signals $\x$ from measurements $\y=\A\x+\v$,
where $\v$ is an $\ell_2$-bounded noise vector (i.e., $\|\v\|_2\leq \epsilon$ for some constant $\epsilon$).
We investigate some sufficient conditions based on the block restricted isometry property (block-RIP)
for exact (when $\v=\0$) and stable (when $\v\neq\0$) recovery of block sparse signals $\x$.
First, on the one hand, we show that if $\A$ satisfies the block-RIP with $\delta_{K+1}<1/\sqrt{K+1}$,
then every block $K$-sparse signal $\x$ can be exactly or stably recovered by BOMP in $K$ iterations.
On the other hand, we show that, for any $K\geq 1$ and $1/\sqrt{K+1}\leq \delta<1$, there exists a matrix $\A$ satisfying the block-RIP with $\delta_{K+1}=\delta$
and a  block $K$-sparse signal $\x$ such that BOMP may fail to recover $\x$ in $K$ iterations.
Then, we study some sufficient conditions for recovering block $\alpha$-strongly-decaying $K$-sparse signals.
We show that if $\A$ satisfies the block-RIP with $\delta_{K+1}<\sqrt{2}/2$,
then every $\alpha$-strongly-decaying
block $K$-sparse signal can be exactly or stably recovered by BOMP in $K$ iterations under some conditions on $\alpha$.
Our newly found sufficient condition on the block-RIP of $\A$ is
less restrictive than that for $\ell_1$ minimization for this special class of sparse signals.
Furthermore, for any $K\geq 1$, $\alpha>1$ and $\sqrt{2}/2\leq \delta<1$, the recovery of $\x$ may fail in $K$ iterations for a sensing matrix
$\A$ which satisfies the block-RIP with $\delta_{K+1}=\delta$.
Finally, we study some sufficient conditions for partial recovery of block sparse signals.
Specifically, if $\A$ satisfies the block-RIP with $\delta_{K+1}<\sqrt{2}/2$,
then BOMP is guaranteed to recover some blocks of $\x$ if these blocks satisfy a sufficient condition.
We further show that this condition is also sharp.

\end{abstract}


\end{frontmatter}

\section{Introduction}
In many applications, such as computer vision, image reconstruction and blind sources separation (see, e.g.,
\cite{CanT05, Don06, CohDD09, HeXDC07, ElaFM10, WriMMSHY10, FR13}),
we need to recover a $K$-sparse signal $\x\in \mathbb{R}^n$ (i.e., $\x$  has at most $K$ nonzero entries) from the following linear model
\beq
\label{e:model}
\y=\A\x+\v,
\eeq
where $\y\in \mathbb{R}^m$ is an observation signal, $\A\in \mathbb{R}^{m\times n}$ ($m\ll n$)
is a known sensing matrix, and $\v \in \mathbb{R}^{m}$ is an $\ell_2$-bounded noise vector
\cite{Fuc05, DonET06, Can08, WenLZ15}
(i.e., $\|\v\|_2\leq \epsilon$ for some constant $\epsilon$).
In addition to the $\ell_2$-bounded noise, there are some other types of noises, such as
the Gaussian noise \cite{CanT07} (i.e., $v\sim \mathcal{N}(0, \sigma^2\I)$)
and the $\ell_{\infty}$-bounded noise \cite{CaiW11} (i.e., $\|\A\v\|_{\infty}\leq \epsilon$
for some constant $\epsilon$).
In this paper, we focus only on the $\ell_2$-bounded noise as our results can be easily
extended to the other two types of noises (see, e.g., \cite{CaiW11}, \cite{WenZWTM17}).

A natural method of recovering $\x$ in \eqref{e:model} is to solve the following $\ell_0$-minimization problem:
\begin{align}
\label{e:l0}
\min\|\x\|_0 :\; \;\text{subject \;to} \;\|\y-\A\x\|_2\leq \epsilon.
\end{align}
However,  \eqref{e:l0} is an NP-hard problem \cite{AmaK09}.
Fortunately, as demonstrated in the pioneering work \cite{CanT05} \cite{Don06b},
under certain conditions,
one can efficiently solve the following $\ell_1$-minimization problem, instead of solving (2),
to stably reconstruct $\x$:
\begin{align}
\label{e:l1}
\min\|\x\|_1 :\; \;\text{subject \;to} \;\|\y-\A\x\|_2\leq \epsilon.
\end{align}

To analyze the recovery performance of sparse recovery algorithms, a commonly used concept is the so-called restricted isometry property (RIP) \cite{CanT05}.
For a sensing matrix $\A\in \mathbb{R}^{m\times n}$ and for any integer $1\leq K\leq n$,
the $K$-restricted isometry constant (RIC) $\delta_K \in (0, 1)$ of $\A$ is the smallest constant such that
\begin{equation}
\label{e:RIP}
(1-\delta_K)\|\x\|_2^2\leq \|\A\x\|_2^2\leq(1+\delta_K)\|\x\|_2^2
\end{equation}
for all $K$-sparse signals $\x$.
Another commonly used concept is the mutual coherence \cite{DonH01} which is defined as
\beq
\mu=\max_{i\neq j}\frac{|\A_i^\top\A_j|}{\|\A_i\|_2\|\A_j\|_2},
\label{mutual}
\eeq
where $\A_i$ ($1\leq i\leq n$) is the $i$-th column of $\A$.

There have been a variety of sufficient conditions for exactly (when $\|\v\|_2 = \0$) or
stably recovering $K$-sparse signals $\x$ (when  $\|\v\|_2 \neq 0$)
by solving \eqref{e:l1}. For example, $\delta_{2K}<\sqrt{2}-1$ \cite{Can08}, $\delta_{2K}<0.4531$
\cite{FouL09}, $\delta_{2K}<0.4652$ \cite{Fou10b} and $\delta_{2K}<\sqrt{2}/2$ \cite{CaiZ13b}.
Moreover, it was shown in \cite{CaiZ13b} and \cite{DavG09} that the exact recovery of $\x$ may not
be possible if $\delta_{2K}\geq\sqrt{2}/2$ which implies that $\delta_{2K}<\sqrt{2}/2$ \cite{CaiZ13b}
is a sharp sufficient condition for stably recovering $K$-sparse signals $\x$ via solving \eqref{e:l1}.

In addition to the $\ell_1$ minimization approach (\ref{e:l1}), the orthogonal matching pursuit
(OMP) algorithm~\cite{TroG07} is another widely-used  algorithm
to solve \eqref{e:l0} due to its efficiency in computation.
Generally speaking, solving \eqref{e:l0} by the OMP algorithm is faster than solving \eqref{e:l1}
by the interior point method or simplex method or their variants.
Thus, the OMP algorithm has attracted much attention.
There are  many sufficient conditions for exactly or stably recovering $\x$ by using the OMP algorithm, see, e.g., \cite{DavW10, LiuT12, MoS12, WanS12, ChaW14}.
Recently, it was shown in \cite{WenZWTM17} \cite{Mo15} that if $\delta_{K+1}<1/\sqrt{K+1}$, the exact recovery of $K$-sparse signals $\x$ can be guaranteed by
using the OMP algorithm in $K$ iterations. Some necessary conditions on the recovery capability of the OMP algorithm can be found in  \cite{MoS12} \cite{WanS12}.
In particular, it was shown in~\cite{WenZL13} that if $\delta_{K+1}\geq 1/\sqrt{K+1}$, the OMP algorithm may fail to
recover a $K$-sparse signal $\x$ in $K$ iterations.

It is easy to see that the sharp sufficient condition $\delta_{K+1}<1/\sqrt{K+1}$ for the OMP algorithm
is much stronger than the condition $\delta_{2K}<\sqrt{2}/2$ for \eqref{e:l1}.
This leads to at least two interesting directions of research.
One is to use more than $K$ iterations in the OMP algorithm to recover a sparse signal
and find a sufficient condition on $\delta_{c(K+1)}$ (for some constant $c$)
to  ensure the exact recovery as discussed in \cite{Zha11, Liv12,Fou12,LivT14,CohDD17,WanS16}.
These results can be described as follows.  When $\delta_{c(K+1)}<\delta^*< 1$ for some constant $\delta^*$,
any $K$-sparse signal can be exactly or stably recovered in $c K$ iterations by the OMP algorithm.
For example, in \cite{Zha11}, $\delta^*=1/3$ with $c=30$. As explained in \cite{CohDD17}, $c$ is dependent on $\delta^*$.
When $\delta^*<1$ is bigger, so is $c$. When $c>1$ is very large, the computational advantage of the OMP will lose.
Another interesting direction is to characterize a class of $K$-sparse signals which can be exactly or stably recovered by the OMP algorithm
in $K$ iterations when $\delta_{K+1}<\sqrt{2}/2$ in order to maintain the computational advantage.
It was shown in \cite{DavW10} that $\delta_{K+1}<1/3$ is a sufficient condition for recovering
$\alpha$-strongly-decaying $K$-sparse signals (see below for the definition of
$\alpha$-strongly-decaying $K$-sparse signal) by using the OMP algorithm in $K$ iterations under
some conditions on $\alpha$ in noise-free case.
This condition was extended to noise-corrupted models in \cite{DinCG13}.
Recently, Xu showed in \cite{Xu15} that under some conditions on $\alpha$, $\delta_{K+1}<\sqrt{2}-1$ is a sufficient condition.
Recently, Herzet \emph{et al.} showed \cite{HerDS16} that under other sufficient condition
on $\alpha$, the mutual coherence $\mu<1/K$ is a sharp condition to
recover $\alpha$-strongly-decaying signals.

Signals with decaying property are frequently encountered in practical applications, such as
in speech communication \cite{HabGC09} and  audio source separation \cite{VinBGB14}.
Without loss of generality, let us assume that all the entries of a $K$-sparse signal $\x$
are ordered as
\[
|x_1|\geq |x_2|\geq\ldots\geq |x_K|\geq 0
\]
and $|x_j|=0$ for $K+1\leq j\leq n$.
Then, a $K$-sparse signal $\x$ is called an $\alpha$-strongly-decaying signal \cite{DavW10} ($\alpha>1$) if
\[
|x_i|\geq \alpha|x_{i+1}|, \quad 1\leq i\leq K-1.
\]
Thus, two questions are naturally raised: (1) what is the RIC based sharp condition of
exactly or stably recovery of $\alpha$-strongly-decaying $K$-sparse signals by using the OMP algorithm
in $K$ iterations under some conditions on $\alpha$?
(2) Is the condition on $\alpha$ sharp?
We shall answer these two questions by Theorems~\ref{t:SDSR} and \ref{t:SDNR}, and
Corollary \ref{c:SDNR2} in Section \ref{s:main} of this paper.

In this paper, we focus on a generalized version of OMP,
the Block OMP (BOMP) algorithm, which was independently proposed in \cite{EldKB10} and \cite{SwiAL09},
respectively.
The main reason that we focus on the BOMP algorithm is that it has better reconstruction properties
than the standard OMP algorithm as shown in \cite{EldKB10}.
Moreover, recovering block-sparse signals are frequently encountered,
such as recovering signals that lie in unions of subspaces \cite{BluD09, EldM09, Eld09,GedE10},
reconstructing multiband signals \cite{MisE09, MisE10},
face recognition \cite{WriYGSM09},
recovering multiple measurement vectors \cite{CotREK05, CheH06, VanF10, LaiL11b} and clustering of data multiple subspaces \cite{ElhV13}.

To define block-sparsity, $\x$ is viewed as a concatenation of blocks.
Similar to \cite{EldKB10}, we assume $\x$ consists of $L$ blocks each
having identical length of $d$ (i.e., $n=Ld$). Then, $\x$ can be
expressed as
\[
\x=[\x^\top[1], \x^\top[2], \ldots, \x^\top[L]]^\top,
\]
where
\[
\x[i]=[x_{d(i-1)+1}, x_{d(i-1)+2}, \ldots, x_{di}]^\top, \quad  1\leq i\leq L.
\]
A signal $\x\in \mathbb{R}^{n}$ is called block $K$-sparse if at most
$K$ blocks $\x[i], i=1, \cdots, K,$ are not zero vectors \cite{EldKB10}.
Mathematically, a signal $\x$ is block $K$-sparse if and only if
\[
\sum_{i=1}^{L}I(\|\x[i]\|_{2}>0)\leq K,
\]
where $I(\cdot)$ is the indicator function (i.e., it takes the value of
$0$ when its argument is zero and $1$ otherwise).
When the block size is 1, block-sparsity reduces to the conventional
sparsity as defined in \cite{CanRT06, Don06}.
In the rest of the paper, the conventional sparsity will be simply
called sparsity, in contrast to block-sparsity.
Similarly, a sensing matrix $\A$ can also be represented as a
concatenation of column blocks, i.e.,
\[
\A=[\A[1], \A[2], \ldots, \A[L]],
\]
where
\[
\A[i]=[\A_{d(i-1)+1}, \A_{d(i-1)+2}, \ldots, \A_{di}], \quad  1\leq i
\leq L.
\]

For any ordered set $S\subset\{1,2,\ldots ,L\}$ with ascending entries,
let $\A[S]$ denote the submatrix of $\A$ that
contains only the blocks indexed by $S$ and $\x[S]$ denote the
subvector of $\x$ that
contains only the blocks indexed by $S$. For instance,
if $S=\{1,3,4\}$, we have $\A[S]=[\A[1],\A[3],\A[4]]$ and
$\x[S]=[\x^\top[1],\x^\top[3],\x^\top[4]]^\top$. Then, the BOMP
algorithm can be formally described below in Algorithm~\ref{a:BOMP}.

\begin{algorithm}[h!]
\caption{The BOMP algorithm (\cite{EldKB10}, \cite{SwiAL09}) }  \label{a:BOMP}
Input: measurements $\y$, sensing matrix $\A$, sparsity $K$ and the number of blocks $L$.\\
Initialize: $k=0, \rr^0=\y, S_0=\emptyset$.\\
Iterate the five steps below until the stopping criterion is met.

\begin{algorithmic}[1]
\STATE $k=k+1$,
\STATE $s^k=\arg \max\limits_{1\leq i\leq L}\|\A[i]^\top\rr^{k-1}\|_2$,
\STATE $S_k=S_{k-1}\bigcup\{s^k\}$,
\STATE $\hat{\x}[S_k]=\arg \min\|\y-\A[S_k]\x\|_2$,
\STATE $\rr^k=\y-\A[S_k]\hat{\x}[S_k]$.
\end{algorithmic}
Output: $\hat{\x}=\arg \min\limits_{\x: \text{supp}(\x)\subset S_K}\|\y-\A\x\|_2$.
\end{algorithm}

To analyze the performance of algorithms for block sparse signals,
the classic RIP was extended to the block-RIP in \cite{EldM09}.
Specifically, $\A$ is said to satisfy the block-RIP with parameter
$\delta_{BK}\in (0, 1)$ if
\begin{equation}
\label{e:BRIP}
(1-\delta_{BK})\|\x\|_2^2\leq \|\A\x\|_2^2\leq(1+\delta_{BK})\|\x\|_2^2
\end{equation}
for all block $K$ sparse signals $\x$, where the smallest constant
$\delta_{BK}$ is called as the block-RIC of $\A$.
For convenience, we simply denote $\delta_{BK}$ by $\delta_{K}$ in
the rest of this paper.
It is worth mentioning that if the entries of $\A$ independently and identically follow
the Gaussian distribution $\mathcal{N}(0, 1/n)$ or the Bernoulli distribution
namely $a_{ij}=\pm1/\sqrt{n}$ with equal probability,
then it is of overwhelming probability that $\A$ satisfies the block-RIP
for certain $m$ and $n$, which respectively denote the row and column numbers of $\A$
(see \cite[Proposition 3]{EldM09} and the second paragraph above \cite[Corollary 3]{EldM09}).

Similarly,  we assume the block entries of
$\x\in \mathbb{R}^{n}$ are ordered as
\beq
\label{e:xbar}
\|\x[1]\|_2\geq \|\x[2]\|_2\geq\ldots\geq \|\x[K]\|_2\geq 0
\eeq
with $\|\x[j]\|_2=\0$ for $K< j\leq L$.
In the following, we define the block $\alpha$-strongly-decaying $K$-sparse signals,
which is an extension of the $\alpha$-strongly-decaying $K$-sparse explained above.

\begin{definition}
\label{d:blockstrong}
Let $\alpha>1$, then a block $K$-sparse signal $\x\in \mathbb{R}^{n}$
is called a block $\alpha$-strongly-decaying $K$-sparse signal if
\beq
\label{e:alphatrong}
\|\x[j]\|_2\geq \alpha\|\x[j+1]\|_2, \quad 1\leq j\leq K-1.
\eeq
\end{definition}

With the above explanation of the concepts and notations, we now
summarize the main contributions of this paper  as follows.
\begin{itemize}
\item
If $\A$ satisfies the block-RIP with  $\delta_{K+1}<1/\sqrt{K+1}$,
then every block $K$-sparse signal can be  exactly or stably recovered
by the BOMP algorithm in $K$ iterations, see Corollaries  \ref{c:SRN} and \ref{c:SR}.
Moreover, for any $K\geq1$ and $1/\sqrt{K+1}\leq \delta<1$, we show that
there exist a matrix $\A$ satisfying the block-RIP with $\delta_{K+1}=\delta$ and a block $K$-sparse signal $\x$
such that the BOMP may fail to recover $\x$ in $K$ iterations, see Theorem~\ref{t:NR}.
These results extend the existing results in \cite{Mo15,WenZWTM17,WenZL13} from $d=1$ to any $d\ge 1$ in a non-trivial way.

\item
If $\A$ satisfies the block-RIP with $\delta_{K+1}<\sqrt{2}/2$, then under some sufficient conditions on $\alpha$,
every block $\alpha$-strongly-decaying $K$-sparse signal can be
exactly  or stably recovered by the BOMP Algorithm in $K$ iterations,
see Theorems~\ref{t:SDSR} and Corollary~\ref{t:SDSRN}.
In addition, for any $K\geq 1$, $\alpha>1$  and $\sqrt{2}/2\leq \delta<1$, we
show that there exist a matrix $\A$ satisfying
the block-RIP with $\delta_{K+1}=\delta$ and an $\alpha$
-strongly-decaying block $K$-sparse signal $\x$
such that the BOMP Algorithm may fail to recover $\x$ in $K$
iterations, see Theorem~\ref{t:SDNR}.
Furthermore, the condition on $\alpha$ is also shown to be tight, see Corollary~\ref{c:SDNR2}.
Obviously, when the block size is 1, then the result for BOMP reduces to that for the
standard OMP. Thus, it improves the result in \cite{DavW10} from $\delta_{K+1}<1/3$ (for OMP)
to $\delta_{K+1}\le \sqrt{2}/2$ (for BOMP).

\item
Finally, we study partial recovery of block $K$-sparse signals.
We show that if $\A$ satisfies the block-RIP with
$\delta_{K+1}<\sqrt{2}/2$ and $\|\x[j]\|_2/\|\x[j+1]\|_2, 1\leq j\leq i$ satisfy a sufficient
condition for some $1\leq i\leq K$ (note that if $i=L$, this class of block $K$-sparse signals
are block $\alpha$-strongly-decaying $K$-sparse signals for certain $\alpha$),
then the BOMP algorithm chooses an index such that the corresponding subvector of $\x$ is a nonzero 
vector in each of the first $i$ iterations,
see Theorem \ref{t:SDSRNP} and Corollary \ref{c:SDSRP}. Moreover, we show that
this condition is tight, see Corollary~\ref{c:SDNRP}.
To the best of our knowledge, partial recovery of sparse signals with BOMP was not
reported before.
\end{itemize}

As far as we know, there is not any reference which studies the recovery of
$\alpha$-strongly-decaying sparse vectors by solving the $\ell_1$-minimization problem \eqref{e:l1}.
Thus, so far, the only known sharp sufficient condition for recovering $\alpha$-strongly-decaying
sparse vectors by solving the $\ell_1$-minimization problem \eqref{e:l1}
is still $\delta_{2K}<\sqrt{2}/2$.
Thus, our sufficient condition $\delta_{K+1}<\sqrt{2}/2$
for using the BOMP algorithm to recover $\alpha$-strongly-decaying sparse vectors is less restrictive.
Also, using the BOMP algorithm, the recovery is done in $K$ iterations which shows its efficiency.

It was shown in \cite{HerDS16} that under a sufficient condition on $\alpha$,
$\mu<1/K$ is a sharp condition for recovering $\alpha$-strongly-decaying sparse vectors with OMP in
$K$ iterations. The condition requires $m\geq \mathcal{O}(K^2\log n)$ observations
(as pointed out in \cite{HerDS16}, this can be seen by invoking the Welch bound
$\mu\geq \sqrt{\frac{n-m}{m(n-1)}}$\cite[Theorem 5.7]{FR13}).
Our sharp condition is $\delta_{K+1}<\sqrt{2}/2$ and only requires $m\geq \mathcal{O}(K\log (n/K))$
(see \cite[Theorem 5.2]{BarDDW08} or \cite[equation (1.3)]{Liv12}),
and hence much less observations are needed.
These further show the effectiveness of the BOMP algorithm.
It is worth mentioning that although finding the mutual coherence of a given $m\times n$ matrix can be
achieved in polynomial time of $n$ according to its definition, it is an NP-hard problem to
find the RIC of a given matrix \cite{BanDMS13} \cite{TilP14}.

The rest of the paper is organized as follows.
In section~\ref{s:pre}, we define some notations and give some useful lemmas that will be needed to prove our main results.
We present our main results in Section~\ref{s:main} and prove them in Section~\ref{s:proofs}.
Finally, we summarize this paper and present some future research problems in Section~\ref{s:con}.

\section{Preliminaries}
\label{s:pre}
In this section, we first introduce some notations which will be used in the rest of this paper.

{\it Notations:} Let $\mathbb{R}$ be the real field. Boldface lowercase letters denote column vectors, and
boldface uppercase letters denote matrices,
e.g., $\x\in\mathbb{R}^n$ and $\A\in\mathbb{R}^{m\times n}$.
For a vector $\x$, $x_{i}$ denotes the $i$-th entry of $\x$.
Let $\e_k$ denote the $k$-th column of an identity matrix $\I$
and $\0$ denote a zero matrix or a zero vector.
Let $\Omega$ be the set of all the indexes $i$ such that $\x[i]$ are not zero vectors for block $K$-sparse signals $\x$,
then $|\Omega|\leq K$, where $|\Omega|$ is the cardinality of $\Omega$.
Let $\Omega \setminus S=\{k|k\in\Omega,k\not\in S\}$ for set $S$.
Let $\Omega^c$ and $S^c$ be the complement of $\Omega$ and $S$,
i.e., $\Omega^c=\{1,2,\ldots ,L\}\setminus \Omega$, and $S^c=\{1,2,\ldots ,L\}\setminus S$.
Let $\A[S]$ be the submatrix of $\A$ that  contains only the blocks of columns indexed by $S$,
and $\x[S]$ be the subvector of $\x$ that  contains only the blocks indexed by $S$,
and $\A^\top[S]$ be the transpose of $\A[S]$.
For any full column rank matrix $\A[S]$, let $\P[S]=\A[S](\A^\top[S]\A[S])^{-1}\A^\top[S]$
and $\P^{\bot}[S]=\I-\P[S]$ denote the projector and orthogonal complement projector
on the column space of $\A[S]$, respectively.

Next we introduce the $\ell_2/\ell_p$-norm for block vectors and some useful lemmas.
\subsection{$\ell_2/\ell_p$-norm}
Like in \cite{EldKB10} and \cite{LaiL11b}, for $\x\in \mathbb{R}^n$, we define a general mixed
$\ell_2/\ell_p$-norm (where $p=1,2,\infty$) as
\beq
\label{e:l2pnorm}
\|\x\|_{2,p}=\|\w\|_p,
\eeq
where $\w\in \mathbb{R}^{L}$ with $w_{\ell}=\|\x[\ell]\|_{2}$ for $1\leq \ell\leq L$.
Note that
$
\|\x\|_{2,2}=\|\x\|_{2}.
$
Thus, we use $\|\x\|_{2}$ instead of $\|\x\|_{2,2}$ for short.
Moreover, if the block size $d$ is equal to 1, for $p=1,2,\infty$, we have
$
\|\x\|_{2,p}=\|\x\|_{p}.
$

\subsection{Some Useful Lemmas}
We can easily get the following lemmas by extending \cite[Lemma 1]{DaiM09},~\cite[Proposition 3.1]{NeeT09} and~\cite[Lemma 1]{SheLPL14}
for $d=1$ to general $d$ (note that $d$ is the length of the blocks).

\begin{lemma}
\label{l:monot}
If $\A$ satisfies the block-RIP of orders $K_1$ and $K_2$ with $K_1<K_2$, then
$
\delta_{K_1}\leq \delta_{K_2}.
$
\end{lemma}

\begin{lemma}
\label{l:AtRIP}
Let $\A$ satisfy the block-RIP of order $K$ and $S$ be a set with $|S|\leq K$. Then
for any $\x \in \mathbb{R}^m$,
\[
\|\A^\top[S]\x\|_2^2\leq(1+\delta_K)\|\x\|_2^2.
\]
\end{lemma}

\begin{lemma}
\label{l:orthogonalcomp}
Let sets $S_1,S_2$ satisfy $|S_2\setminus S_1|\geq 1$ and matrix $\A$ satisfy the block-RIP of order
$|S_1\cup S_2|$. Then for any vector $\x \in \mathbb{R}^{|S_2\setminus S_1|\times d}$,
\begin{align*}
(1-\delta_{|S_1\cup S_2|})\|\x\|_2^2 &\leq \|\P^{\bot}[S_1]\A[S_2\setminus S_1]\x\|_2^2
\leq(1+\delta_{|S_1\cup S_2|})\|\x\|_2^2.
\end{align*}
\end{lemma}

\section{Main Results}
\label{s:main}

\subsection{An All-Purpose Sufficient Condition}
In this subsection, we give an all-purpose sufficient condition for the recovery of
block $K$-sparse signals with the BOMP algorithm.
We start with the following useful lemma.
\begin{lemma}
\label{l:main}
Let $\A$ in~\eqref{e:model} satisfy the block-RIP of order $K+1$ and $S$ be a proper subset of $\Omega\neq\emptyset$.
If $\x$ in~\eqref{e:model} satisfies
\begin{align}
\label{e:l1l2inequality}
\|\x[\Omega\setminus S]\|_{2,1}^2\leq \nu \|\x[\Omega\setminus S]\|_{2}^2
\end{align}
for some $\nu\geq 1$, then
\begin{align}
\label{e:main}
\|\A^\top[\Omega\setminus S]\|_{2,\infty}-\|\A^\top[\Omega^c]\q[S]\|_{2,\infty}
\geq  \frac{(1-\sqrt {\nu+1 }\delta_{|\Omega|+1})\|\x[\Omega\setminus S]\|_2}{\sqrt{\nu}},
\end{align}
where
\beq
\label{e:qs}
\q[S]=\P^{\bot}[S]\A[\Omega\setminus S]\x[\Omega\setminus S].
\eeq
\end{lemma}

\begin{proof}
See Section~\ref{ss:profLmain}.
\end{proof}

By using Lemma \ref{l:main}, we can prove the following all-purpose sufficient condition
for the stable recovery of block $K$-sparse signals with the BOMP algorithm.
\begin{theorem}
\label{t:SDG}
Let $f(t)$ be a nondecreasing function of $t>0$ with $0<f(t)\leq t$.
Suppose that block $K$-sparse $\x$  satisfies
\begin{align}
\label{e:l1l2inequalityG}
\|\x[\Omega\setminus S]\|_{2,1}^2\leq f(|\Omega\setminus S|) \|\x[\Omega\setminus S]\|_{2}^2
\end{align}
for any $S$ which is a proper subset of $\Omega\neq\emptyset$.
Suppose that $\v$ satisfies $\|\v\|_2\leq \epsilon$ and $\A$ satisfies the block-RIP of order $K+1$ with
\beq
\label{e:deltaG}
\delta_{K+1} < \frac{1}{\sqrt {f(K)+1 }}.
\eeq
Then, the BOMP algorithm with the stopping criterion $\|\rr^k\|\leq \epsilon$
exactly recovers $\Omega$ in $K$ iterations provided that
\beq
\label{e:suffgenenoise}
\min_{i\in\Omega}\|\x[i]\|_2>\frac{2\epsilon}{1-\sqrt {f(K)+1 }\delta_{K+1}}.
\eeq
Moreover, the recovery error is bounded by
\beq
\label{e:xerr}
\|\x-\hat{\x}\|_2\leq \epsilon,
\eeq
where $\hat{\x}$ is the block $K$-sparse signal returned by the BOMP algorithm.
\end{theorem}

\begin{proof}
See Section~\ref{ss:profTsuffgeneral}.
\end{proof}

\begin{remark}
By \eqref{e:l2pnorm} and the Cauchy-Schwarz inequality, it is easy to see that \eqref{e:l1l2inequalityG} always
holds if $f(t)=t$.
So it is not necessary to consider the case that $f(t)>t$ and this is the reason why we assume $f(t)\leq t$.
\end{remark}

\subsection{Recovery of General Block Sparse Signals}
In this subsection, we study the recovery of general block sparse signals.
On the one hand, we will show that if $\A$  satisfies the block-RIP of order $K+1$ with
$\delta_{K+1} < 1/\sqrt {K+1 }$,
then the BOMP Algorithm exactly or stably (under some conditions on $\min_{i\in\Omega}\|\x[i]\|_2$)
recovers block $K$-sparse signals $\x$ in $K$ iterations.
On the other hand, we will show that if $\delta_{K+1} \geq 1/\sqrt {K+1 }$, then
the BOMP algorithm may fail to recover a block $K$-sparse signal $\x$ in $K$ iterations.

By the Cauchy-Schwarz inequality and \eqref{e:l2pnorm},
it is easy to see that \eqref{e:l1l2inequalityG} holds if $f(t)=t$.
Consequently, by Theorem~\ref{t:SDG}, we can immediately obtain
the following result which provides a sufficient condition for
the exact support recovery of block $K$-sparse signal $\x$ with BOMP in $K$ iterations
in noisy case.

\begin{corollary}
\label{c:SRN}
Suppose that $\v$ in~\eqref{e:model} satisfies $\|\v\|_2\leq \epsilon$,
and $\A$ in~\eqref{e:model} satisfies the  block-RIP of order $K+1$ with
\beq
\label{e:deltaSR}
\delta_{K+1} < \frac{1}{\sqrt {K+1 }}.
\eeq
Then the BOMP algorithm with the stopping criterion $\|\rr^k\|\leq \epsilon$ exactly recovers
$\Omega$ in $K$ iterations provided that
\[
\min_{i\in\Omega}\|\x[i]\|_2>\frac{2\epsilon}{1-\sqrt {K+1 }\delta_{K+1}}.
\]
Moreover, the recovery error can be bounded by \eqref{e:xerr}.
\end{corollary}

From Corollary \ref{c:SRN}, we can immediately obtain the following sufficient condition for
the exact recovery of block $K$-sparse signals with BOMP in $K$ iterations in noise-free case.

\begin{corollary}
\label{c:SR}
Suppose that $\A$ in~\eqref{e:model} satisfies the block-RIP of order $K+1$ with \eqref{e:deltaSR}.
Then, the BOMP algorithm exactly recovers block $K$-sparse signals $\x$ in $K$ iterations when $\v=\0$.
\end{corollary}

When the block size $d=1$, the block-RIP reduces to the classic RIP and the BOMP algorithm reduces
to the OMP algorithm.
Then Corollary \ref{c:SRN} and Corollary \ref{c:SR} are \cite[Theorem 1]{WenZWTM17}
and \cite[Theorem III.1]{Mo15}, respectively.

In the following, we give a necessary condition in terms of the RIC for the exact recovery of $K$-block
sparse signals with the BOMP algorithm.
\begin{theorem}
\label{t:NR}
For any given positive integers $d, K\geq 1$ and any
\beq
\label{e:deltaNR}
\frac{1}{\sqrt{K+1}}\leq \delta<1,
\eeq
there always exist a block $K$-sparse signal $\x$ and
a matrix $\A$ satisfying the block-RIP of order $K+1$ with $\delta_{K+1}=\delta$
such that the BOMP algorithm may fail to recover $\x$ in $K$ iterations.
\end{theorem}
\begin{proof}
See Section~\ref{ss:pfTNR}.
\end{proof}

When the block size $d=1$, Theorem~\ref{t:NR} reduces to \cite[Theorem 1]{WenZL13}.
Moreover, from Corollary \ref{c:SR} and  Theorem \ref{t:NR}, we can see that \eqref{e:deltaSR}
is a sharp sufficient condition for recovering block $K$-sparse signals by the BOMP algorithm
in $K$ iterations.

\begin{remark}
Note that the sharp sufficient condition \eqref{e:deltaSR} has been independently obtained
in \cite{CheG17} (see \cite[Theorems 3.3, 3.4, 3.5]{CheG17})
which studies the recovery of block $K$-sparse signals
with block orthogonal multi-matching pursuit, an improvement of BOMP.
\end{remark}

\subsection{Recovery of  Strongly-Decaying Block Sparse Signals}
In this subsection, we will show that if $\A$  satisfies the block-RIP of order $K+1$
with $\delta_{K+1} < \sqrt{2}/2$,
then the BOMP algorithm exactly or stably (under some constraints on $\min_{i\in\Omega}\|\x[i]\|_2$)
recovers all  block $\alpha$-strongly-decaying $K$-sparse signals  in $K$ iterations
if $\alpha$ is larger than a value related to $\delta_{K+1}$ and $K$.
Furthermore, we will show that if $\delta_{K+1} \geq \sqrt{2}/2$,
the BOMP algorithm may fail to recover a block $\alpha$-strongly-decaying $K$-sparse signal
$\x$ in $K$ iterations,  regardless how large is $\alpha$.
We will also show that the condition on $\alpha$  is sharp.

We first introduce the following lemma.
\begin{lemma}
\label{l:gi}
For each $1\leq i\leq K$, let
\beq
\label{e:gi}
g_i(t)=\frac{(t^i-1)(t+1)}{(t^i+1)(t-1)}, \quad t>1.
\eeq
Then
\beq
\label{e:giorder}
1=g_1(t)< g_2(t)<\ldots < g_K(t).
\eeq
Moreover, $g_i(t)$ is strictly decreasing with $t$ and  $1< g_i(t)<i$ for $2< i\leq K$.
\end{lemma}

\begin{proof}
See Section~\ref{ss:pfLg}.
\end{proof}

By Lemma \ref{l:gi}, for $1<s< K$, $g_K(t)=s$ has a unique solution $t_s$.
To simplify the notation, we define
\beq
\label{e:ginverse}
g_K^{-1}(s)=
\begin{cases}
t_s, & 1<s< K \\
1, & s\geq K
\end{cases}.
\eeq
Note that $g_K^{-1}(s)$ can be easily computed, for example, using Newton's method.

By \eqref{e:ginverse} and Theorem \ref{t:SDG}, we can obtain the following result
which provides a sufficient condition for the exact support recovery of
block $\alpha$-strongly-decaying $K$-sparse signals with BOMP in $K$ iterations
in the noisy case.

\begin{theorem}
\label{t:SDSRN}
Suppose that $\v$ in~\eqref{e:model} satisfies $\|\v\|_2\leq \epsilon$,
and $\A$ in~\eqref{e:model} satisfies the block-RIP of order $K+1$ with
\beq
\label{e:deltaSDSR}
\delta_{K+1} < \frac{\sqrt {2}}{2}.
\eeq
If $\x$ is a block $\alpha$-strongly-decaying $K$-sparse signal with $\alpha$ satisfying \beq
\label{e:alphaCSD}
\alpha> g_{K}^{-1}(\delta_{K+1}^{-2}-1).
\eeq
Then BOMP with the stopping criterion $\|\rr^k\|\leq \epsilon$ exactly recovers $\Omega$
in $K$ iterations provided that
\begin{align*}
\min_{i\in\Omega}\|\x[i]\|_2>\frac{2\epsilon}{1-\sqrt {1+\min\{s^{-2}-1,K\} }\delta_{K+1}},
\end{align*}
where $s$ satisfies
\beq
\label{e:s}
\delta_{K+1}<s< \frac{\sqrt {2}}{2},\,\;g_{K}^{-1}(\delta_{K+1} ^{-2}-1)<g_{K}^{-1}(s^{-2}-1)\leq\alpha.
\eeq
Moreover, the recovery error can be bounded by \eqref{e:xerr}.
\end{theorem}

\begin{proof}
See Section~\ref{SS:pfTSDSRN}.
\end{proof}

By Theorem \ref{t:SDSRN}, one can easily obtain the following result which
provides a sufficient condition for the exact recovery of
block $\alpha$-strongly-decaying $K$-sparse signals with BOMP in $K$ iterations
in the noise-free case.

\begin{corollary}
\label{t:SDSR}
Suppose that $\A$ in~\eqref{e:model} satisfies the block-RIP of order $K+1$ with \eqref{e:deltaSDSR},
and $\x$ is a block $\alpha$-strongly-decaying $K$-sparse signal with \eqref{e:alphaCSD}.
Then the BOMP algorithm exactly recovers $\x$ in $K$ iterations when $\v=\0$.
\end{corollary}

\begin{remark}
By \eqref{e:ginverse}, $g_{K}^{-1}(\delta_{K+1} ^{-2}-1)$ is well defined for $\delta_{K+1} <\sqrt {2}/2$.
By Lemma \ref{l:gi}, $g_K(t)$ is strictly decreasing with $t$, so $g_K^{-1}(s)$ is strictly decreasing with $s$.
Therefore, for each $\alpha>g_{K}^{-1}(\delta_{K+1} ^{-2}-1)$, there exists $s$ such that \eqref{e:s} holds.
\end{remark}

It was shown in \cite[Theorem 4.1]{DavW10} that if $\A$ in~\eqref{e:model} satisfies the RIP of order $K+1$ with
$\delta_{K+1}<1/3$, and $\x$ is an $\alpha$-strongly-decaying $K$-sparse signal with
\beq
\label{e:alphaCSDE}
\alpha> \frac{1+2\frac{\delta_{K+1}}{1-\delta_{K+1}}\sqrt{K-1}}{1-2\frac{\delta_{K+1}}{1-\delta_{K+1}}}.
\eeq
Then, the OMP algorithm exactly recovers $\x$ in $K$ iterations when $\v=\0$.

If the block size $d=1$, then the sufficient condition given by Theorem \ref{t:SDSR} is a condition
for recovering $\alpha$-strongly-decaying $K$-sparse signals by using the OMP algorithm.
By \eqref{e:deltaSDSR}, our condition on $\delta_{K+1}$ is much weaker than that given in
\cite[Theorem 4.1]{DavW10}.
Moreover, our condition on $\alpha$ is also less restrictive than that given in
\cite[Theorem 4.1]{DavW10}.
In the following, we show it. By \eqref{e:alphaCSD} and \eqref{e:alphaCSDE},
it is equivalent to show that
\[
 \frac{1+2\frac{\delta_{K+1}}{1-\delta_{K+1}}\sqrt{K-1}}{1-2\frac{\delta_{K+1}}{1-\delta_{K+1}}}
 > g_{K}^{-1}(\delta_{K+1}^{-2}-1)
\]
which is equivalent to
\[
 \frac{1-\delta_{K+1}+2\delta_{K+1}\sqrt{K-1}}{1-3\delta_{K+1}}
 > g_{K}^{-1}(\delta_{K+1}^{-2}-1).
\]
By the previous analysis, $g_{K}^{-1}(s)$ is decreasing with $s$. Thus, we only need to show
\[
g_{K}\left( \frac{1-\delta_{K+1}+2\delta_{K+1}\sqrt{K-1}}{1-3\delta_{K+1}}\right)
< \delta_{K+1}^{-2}-1.
\]
By \eqref{e:gi}, the inequality above holds if
\[
\frac{\frac{1-\delta_{K+1}+2\delta_{K+1}\sqrt{K-1}}{1-3\delta_{K+1}}+1}
{\frac{1-\delta_{K+1}+2\delta_{K+1}\sqrt{K-1}}{1-3\delta_{K+1}}-1}
< \delta_{K+1}^{-2}-1,
\]
i.e.,
\[
\frac{1-2\delta_{K+1}+\delta_{K+1}\sqrt{K-1}}{\delta_{K+1}(1+\sqrt{K-1})}
< \frac{1-\delta_{K+1}^{2}}{\delta_{K+1}^{2}}.
\]
To show the aforementioned inequality, it is equivalently to show
\[
2\delta_{K+1}^2\sqrt{K-1}<(1+\delta_{K+1}^2-\delta_{K+1})+\sqrt{K-1}.
\]
It is clear to see that the above inequality holds when $\delta_{K+1}<\sqrt{2}/2$.
Thus, our sufficient condition given by Theorem \ref{t:SDSR} is less restrictive than
\cite[Theorem 4.1]{DavW10} in terms of both $\delta_{K+1}$ and $\alpha$.

In the following, we give two necessary conditions in terms of the RIC and $\alpha$
for the exact recovery of block $\alpha$-strongly-decaying $K$-sparse signals
by using the BOMP algorithm.
\begin{theorem}
\label{t:SDNR}
For any given positive integers $d, K\geq 1$, any given $\alpha>1$ and any
\beq
\label{e:deltaSDNR}
\frac{\sqrt {2}}{2}\leq \delta<1,
\eeq
there always exist a block $\alpha$-strongly-decaying $K$-sparse signal $\x$
and a matrix $\A$ satisfying the block-RIP of order $K+1$ with $\delta_{K+1}=\delta$
such that the BOMP algorithm may fail to recover $\x$ in $K$ iterations.
\end{theorem}
\begin{proof}
See Section~\ref{SS:pfTSDNR}.
\end{proof}

Let
\[
\delta_{K+1}=\frac{1}{\sqrt{K+1}}.
\]
By \eqref{e:ginverse}, we have
\[
g_{K}^{-1}(\delta_{K+1}^{-2}-1)=g_{K}^{-1}(K)=1.
\]
By Definition \ref{d:blockstrong}, $\alpha>1$, thus we cannot set $\alpha=g_{K}^{-1}(\delta_{K+1}^{-2}-1)$
for all $\delta_{K+1}$ satisfying \eqref{e:deltaSDSR}. Hence, we have the following result.

\begin{corollary}
\label{c:SDNR2}
For any given positive integers $d, K\geq 1$, there always exist a matrix $\A$ which satisfies
the block-RIP of order $K+1$ with $\delta_{K+1}$ satisfying \eqref{e:deltaSDSR} such that
\[
\alpha= g_{K}^{-1}(\delta_{K+1}^{-2}-1)=1,
\]
where $g_{K}(t)$ is defined in \eqref{e:gi}.
\end{corollary}

From Theorems \ref{t:SDSRN}, \ref{t:SDNR} and Corollary \ref{c:SDNR2}, we can see that \eqref{e:deltaSDSR}
is a sharp sufficient condition for recovering block $\alpha$-strongly-decaying $K$-sparse signals $\x$
by the BOMP algorithm in $K$ iterations.

\subsection{Partial Recovery of  Block Sparse Signals}
In this subsection, we will show that if $\A$ satisfies the block-RIP of order $K+1$
with $\delta_{K+1}<\sqrt{2}/2$
and $\|\x[j]\|_2/\|\x[j+1]\|_2, 1\leq j\leq i$ satisfy some conditions for some $1\leq i\leq K$,
then the BOMP algorithm is able to choose an index in $\Omega$ in each of the first $i$ iterations.
Moreover, we will show that this condition is sharp.

Before presenting our  sufficient conditions, we introduce the following lemma.
\begin{lemma}
\label{l:hat}
For each $1\leq i\leq K-1$, let
\beq
\label{e:hat}
\hat{h}_i(t)= \frac{(t+i)^2}{t^2+i}, \quad t\geq1.
\eeq
Then,  $\hat{h}_i$ is strictly decreasing with $t$, and $1<\hat{h}_i(t)\leq 1+i$.

Moreover,  for $1<s\leq 1+i$, $\hat{h}_i(t)=s$ if and only if
$t=h_i(s)$, where for $1\leq i\leq K-1$,
\beq
\label{e:hi}
h_i(s)=\frac{i+\sqrt{i^2s+(s-s^2)i}}{s-1}, \,\; 1< s\leq 1+i.
\eeq
\end{lemma}

Note that for each $1\leq i\leq K-1$ and $1< s\leq 1+i$, we have
$i^2s+(s-s^2)i=is[(i+1)-s]\geq0.$ Thus, \eqref{e:hi} is well-defined.

{\em Proof}. By some simple calculations and the assumption that $t\geq1$, we have
\[
\hat{h}_i(t)'= \frac{2(t+i)(1-t)i}{(t^2+i)^2}\leq0.
\]
Moreover, $\hat{h}_i(t)'=0$ if and only if $t=1$.
Thus, $\hat{h}_i$ is strictly decreasing with $t$. It is easy to see that
\[
\hat{h}_i(1)=1+i, \quad \lim_{t\rightarrow+\infty}\hat{h}_i(t)=1.
\]
Thus, $1<\hat{h}_i(t)\leq 1+i$. For $1<s\leq 1+i$, it is straightforward to solve the quadratic equation
$\hat{h}(t)=s$ which yields \eqref{e:hi}.  $\Box$

By \eqref{e:hi} and Theorem \ref{t:SDG},
we can obtain the following sufficient condition for partial support recovery of block $K$-sparse
signals with the BOMP algorithm in the noisy case.

\begin{theorem}
\label{t:SDSRNP}
Suppose that in~\eqref{e:model}, $\A$  satisfies the block-RIP of order $K+1$ with $\delta_{K+1}$
satisfying \eqref{e:deltaSDSR} and $\v$ satisfies $\|\v\|_2\leq \epsilon$.
If the block $K$-sparse $\x$  in~\eqref{e:model} satisfies
\beq
\label{e:alphaCSDP}
\|\x[j]\|_2> h_{K-j}(\delta_{K+1} ^{-2}-1)\|\x[j+1]\|_2, \quad 1\leq j\leq i.
\eeq
Then the BOMP algorithm choose  an index in $\Omega$ in each of the first $i$ iterations
provided that
\begin{align*}
\min_{i\in\Omega}\|\x[i]\|_2>\frac{2\epsilon}{1-\sqrt {1+\min\{s^{-2}-1,K\} }\delta_{K+1}},
\end{align*}
where $s$ satisfies $\delta_{K+1}< s < \frac{\sqrt {2}}{2}$ and
\beq
\label{e:spartial}
h_{K-1}(\delta_{K+1} ^{-2}-1)<h_{K-1}(s^{-2}-1)\leq\min_{1\leq j\leq i}\frac{\|\x[j]\|_2}{\|\x[j+1]\|_2}.
\eeq
\end{theorem}

\begin{proof}
See Section~\ref{SS:pfTSDSRNP}.
\end{proof}

By Theorem \ref{t:SDSRNP}, we can get the following consequence which provides a sufficient condition
for the partial recovery of block $K$-sparse signals with the BOMP algorithm in the noise-free case.
\begin{corollary}
\label{c:SDSRP}
Suppose that $\A$ in~\eqref{e:model} satisfies the block-RIP of order $K+1$ with $\delta_{K+1}$
satisfying \eqref{e:deltaSDSR}. If  the block $K$-sparse $\x$  in~\eqref{e:model} satisfies
\eqref{e:alphaCSDP}.
Then, the BOMP algorithm choose an index in $\Omega$ in each of the first $i$ iterations when $\v=\0$.
\end{corollary}

\begin{remark}
By Lemma \ref{l:hat}, $\hat{h}_1(t)$ is strictly decreasing with $t$, so $h_1(s)$ is strictly decreasing
with $s$. Therefore, by \eqref{e:alphaCSDP}, there exists an $s$ such that \eqref{e:spartial} holds.
\end{remark}

In the following, we give a necessary condition for partial recovery of block $K$-sparse signals
by using the BOMP algorithm.  Let
\[
\delta_{K+1}=\frac{1}{\sqrt{K+1}}.
\]
By \eqref{e:hi}, we have
\[
h_{K-1}(\delta_{K+1} ^{-2}-1)=h_{K-1}(K)=1.
\]
Thus, from the proof of Theorem \ref{t:NR}, we can easily get the following result.
\begin{corollary}
\label{c:SDNRP}
For any given positive integers $d, K\geq 1$, there always exist a matrix $\A$ satisfying the
block-RIP of order $K+1$ with $\delta_{K+1}$ satisfying \eqref{e:deltaSDSR},
and  a block $K$-sparse signal with
\beq
\label{e:alphaCSNDP}
\|\x[1]\|_2= h_{K-1}(\delta_{K+1} ^{-2}-1)\|\x[2]\|_2
\eeq
such that the BOMP  algorithm may fail to choose an index in the support $\Omega$
in the first iteration.
\end{corollary}

From Corollaries \ref{c:SDSRP} and \ref{c:SDNRP}, and Theorem \ref{t:SDSRNP},
we can see that Corollary \ref{c:SDSRP}
gives a sharp sufficient condition for partial recovery of block $K$-sparse signals $\x$
with the BOMP algorithm in $K$ iterations.

\section{Proofs}
\label{s:proofs}
In this section, we mainly prove the main results given in Section~\ref{s:main}. We devote each subsection
to one proof.

\subsection{Proof of Lemma~\ref{l:main}}
\label{ss:profLmain}
The following proof follows a similar line as that for \cite[Lemma 1]{WenZWTM17}, but with a key distinction.
Specifically, we introduce a well-chosen vector $\h$ (see \eqref{e:h}) to define the vector $\w$ (see \eqref{e:w})
which incorporates the particular feature of block sparse.

{\em Proof}. The proof consists of two steps, in the first step, we prove
\begin{align}
\label{e:main1}
\sqrt{\nu}\|\x[\Omega\setminus S]\|_{2}\|\A^\top[\Omega\setminus S]\q[S]\|_{2, \infty}\geq\|\q[S]\|_2^2.
\end{align}
In the second step, we show for each $j\in \Omega^c$,
\begin{align}
\label{e:main2}
\|\q[S]\|_2^2-&\;\,\sqrt{\nu}\|\x[\Omega\setminus S]\|_{2}\|\A^\top[j]\q[S]\|_2
\geq\;\,(1-\sqrt {\nu+1 }\delta_{|\Omega|+1})\|\x[\Omega\setminus S]\|_2^2.
\end{align}

We first show~\eqref{e:main1}. It is easy to check that
\begin{align*}
\quad&\sqrt{\nu}\|\x[\Omega\setminus S]\|_{2}\|\A^\top[\Omega\setminus S]\q[S]\|_{2,\infty}
\overset{(a) }{\geq}\|\x[\Omega\setminus S]\|_{2,1}\|\A^\top[\Omega\setminus S]\q[S]\|_{2,\infty}\\
\overset{(b)}{=}&\big(\sum_{\ell\in \Omega\setminus S}\|\x[\ell]\|_2\big)\|\A^\top[\Omega\setminus S]\q[S]\|_{2,\infty}
\overset{(c)}{\geq}\sum_{\ell\in \Omega\setminus S}(\|\x[\ell]\|_2\|\A^\top[\ell]\q[S]\|_{2})\\
\overset{(d)}{\geq}&\sum_{\ell\in \Omega\setminus S}(|\x^\top[\ell]\A^\top[\ell]\q[S]|)
\geq\sum_{\ell\in \Omega\setminus S}(\x^\top[\ell]\A^\top[\ell])\q[S]=(\A[\Omega\setminus S]\x[\Omega\setminus S])^\top\q[s]\\
\overset{(e)}{=}&(\P^{\bot}[S]\A[\Omega\setminus S]\x[\Omega\setminus S])^\top(\P^{\bot}[S]\A[\Omega\setminus S]\x[\Omega\setminus S])
=\|\P^{\bot}[S]\A[\Omega\setminus S]\x[\Omega\setminus S]\|_2^2\overset{(f)}{=}\|\q[S]\|_2^2,
\end{align*}
where (a) and (b) respectively follow from \eqref{e:l1l2inequality} and \eqref{e:l2pnorm},
(c) is because for each $\ell\in \Omega\setminus S$,
\[
\|\A^\top[\Omega\setminus S]\q[S]\|_{2,\infty}\geq \|\A[\ell]^\top\q[S]\|_2,
\]
(d) is from the Cauchy-Schwarz inequality, and (e) and (f) are from \eqref{e:qs} and the projection property:
\begin{align}
\label{e:orthcom}
(\P^{\bot}[S])^\top
\P^{\bot}[S]=\P^{\bot}[S]\P^{\bot}[S]=\P^{\bot}[S].
\end{align}
Thus, \eqref{e:main1} holds.

In the following, we prove \eqref{e:main2}.
Let
\[
\mu=-\frac{\sqrt{\nu+1}-1}{\sqrt{\nu}}.
\]
By a simple calculation, we have
\beq
\label{e:muproperty}
\frac{2\mu}{1-\mu^2}=-\sqrt{\nu}, \quad \frac{1+\mu^2}{1-\mu^2}=\sqrt{\nu+1}.
\eeq

To simplify notation, we define $\h \in \mathbb{R}^{d}$ by
\beq
\label{e:h}
h_i=\frac{(\A[j])_{i}^\top \q[S]}{\|\A^\top[j]\q[S]\|_2}, \quad 1\leq i\leq d.
\eeq
Then, it is easy to see that $\|\h\|_2=1$ and
\begin{align}
\label{e:vectorh}
\h^\top\A^\top[j]\q[S]=\|\A^\top[j]\q[S]\|_2.
\end{align}
Furthermore, we define two vectors
\begin{align}
\u=& \bmx \x[\Omega\setminus S]\\ \0
\emx\in \mathbb{R}^{(|\Omega\setminus S|+1)d}, \nonumber\\
\w=& \bmx
\0\\\mu \|\x[\Omega\setminus S]\|_2\h
\emx\in \mathbb{R}^{(|\Omega\setminus S|+1)d}.
\label{e:w}
\end{align}
For given $j\in \Omega^c$, we let
\begin{align}
\label{e:B}
\B=&\P^{\bot}[S]
\bmx
\A[\Omega\setminus S]&\A[ j]
\emx.
\end{align}
Then
\begin{align}
\label{e:AB}
\q[S]=\P^{\bot}[S]\A[\Omega\setminus S]\x[\Omega\setminus S]=\B\u
\end{align}
and
\begin{align}
\label{e:uw}
\|\u+\w\|_2^2&=(1+\mu^2)\|\x[\Omega\setminus S]\|_2^2, \\
\label{e:alphauw}
\|\mu^2\u-\w\|_2^2&=\mu^2(1+\mu^2)\|\x[\Omega\setminus S]\|_2^2.
\end{align}
Moreover, we have
\begin{align*}
\w^\top\B^\top\B\u
\overset{(a)}{=}&\mu\|\x[\Omega\setminus S]\|_2\h^\top\A^\top[j](\P^{\bot}[S])^\top\q[S]\\
\overset{(b)}{=}&\mu\|\x[\Omega\setminus S]\|_2\h^\top \A^\top[j]\q[S]\\
\overset{(c)}{=}&\mu\|\x[\Omega\setminus S]\|_2\|\A^\top[j]\q[S]\|_2,
\end{align*}
where (a) follows from~\eqref{e:B} and~\eqref{e:AB},
(b) is from \eqref{e:qs} and~\eqref{e:orthcom},
and (c) follows from~\eqref{e:vectorh}.
Therefore,
\begin{align}
&\|\B(\u+\w)\|_2^2-\|\B(\mu^2\u-\w)\|_2^2 \nonumber\\
=&(1-\mu^4)\|\B\u\|_2^2+2(1+\mu^2)\w^\top\B^\top\B\u \nonumber\\
=&(1-\mu^4)\|\B\u\|_2^2+2\mu(1+\mu^2)\|\x[\Omega\setminus S]\|_2\|\A^\top[j]\q[S]\|_2  \nonumber\\
=&(1-\mu^4)(\|\B\u\|_2^2+\frac{2\mu}{1-\mu^2}\|\x[\Omega\setminus S]\|_2\|\A^\top[j]\q[S]\|_2) \nonumber\\
=&(1-\mu^4)(\|\B\u\|_2^2-\sqrt{\nu}\|\x[\Omega\setminus S]\|_2\|\A^\top[j]\q[S]\|_2),
\label{e:transf11}
\end{align}
where the last equality follows from the first equality in~\eqref{e:muproperty}.

On the other hand, using \eqref{e:B} and Lemma~\ref{l:orthogonalcomp}, we have
\begin{align}
\,\;&\|\B(\u+\w)\|_2^2-\|\B(\mu^2\u-\w)\|_2^2\nonumber \\
\overset{(a)}{\geq}&(1-\delta_{|\Omega|+1})\|(\u+\w)\|_2^2
-(1+\delta_{|\Omega|+1})\|(\mu^2\u-\w)\|_2^2 \nonumber \\
\overset{(b)}{=}&(1-\delta_{|\Omega|+1})(1+\mu^2)\|\x[\Omega\setminus S]\|_2^2
-(1+\delta_{|\Omega|+1})\mu^2(1+\mu^2)\|\x[\Omega\setminus S]\|_2^2\nonumber \\
=&(1+\mu^2)\|\x[\Omega\setminus S]\|_2^2\big((1-\delta_{|\Omega|+1})-(1+\delta_{|\Omega|+1})\mu^2\big)
\nonumber \\
=&(1+\mu^2)\|\x[\Omega\setminus S]\|_2^2\big((1-\mu^2)-\delta_{|\Omega|+1}(1+\mu^2)\big)\nonumber \\
=&(1-\mu^4)\|\x[\Omega\setminus S]\|_2^2\big(1-\frac{1+\mu^2}{1-\mu^2}\delta_{|\Omega|+1}\big)\nonumber \\
\overset{(c)}{=}&(1-\mu^4)\|\x[\Omega\setminus S]\|_2^2\big(1-\sqrt{\nu+1}\delta_{|\Omega|+1}\big),
\label{e:transf12}
\end{align}
where (a) follows from Lemma~\ref{l:orthogonalcomp} and~\eqref{e:B},
(b) is from \eqref{e:uw} and~\eqref{e:alphauw},
and (c) follows from the second equality in~\eqref{e:muproperty}.

By~\eqref{e:transf11},~\eqref{e:transf12} and the fact that $1-\mu^4>0$, we have
\begin{align*}
\|\B\u\|_2^2-\sqrt{\nu}\|\x[\Omega\setminus S]\|_2\|\A^\top[j]\q[S]\|_2
\geq \|\x[\Omega\setminus S]\|_2^2\big(1-\sqrt{\nu+1}\delta_{|\Omega|+1}\big).
\end{align*}
By \eqref{e:AB}, \eqref{e:main2} holds, and this completes the proof.
 \ \ $\Box$

\subsection{Proof of Theorem~\ref{t:SDG}}
\label{ss:profTsuffgeneral}
{\em Proof}. We first prove the first part of the results in Theorem~\ref{t:SDG}.
The proof consists of two steps.
In the first step, we show that the BOMP algorithm selects correct indexes in all iterations.
In the second step, we prove that it performs exactly $|\Omega|$ iterations.

We prove the first step by induction. Suppose that the BOMP algorithm selects correct indexes
in the first $k$ iterations, i.e., $S_{k}\subseteq \Omega$ for $0\leq k\leq |\Omega|-1$.
Then, we need to show that the BOMP algorithm selects a correct index in the $(k+1)$-th iteration, i.e., by Algorithm~\ref{a:BOMP}, we show that $s^{k+1}\in \Omega$.
Note that the proof for the first selection is contained in the case that $k=0$
because the induction assumption $S_{k}\subseteq \Omega$ holds in this case (not that $S_0=\emptyset$).

By Algorithm \ref{a:BOMP}, it is easy to see that
\[
\|\A^\top[S_{k}]\rr^{k}]\|_{2}=0.
\]
Thus, by line 2 of Algorithm~\ref{a:BOMP}, to show $s^{k+1}\in \Omega$, it suffices to show
\begin{align}
\label{e:generalcond}
\|\A^\top[\Omega\setminus S_{k}]\rr^{k}]\|_{2,\infty}>\|\A^\top[\Omega^c]\rr^{k}]\|_{2,\infty}.
\end{align}

By line 4 of Algorithm~\ref{a:BOMP}, we have
\begin{align}
\label{e:xtk-1}
\hat{\x}[S_{k}]=(\A^\top[S_{k}]\A[S_{k}])^{-1}\A^\top[S_{k}]\y.
\end{align}
Then, by line 5 of Algorithm~\ref{a:BOMP} and~\eqref{e:xtk-1}, we have
\begin{align}
\label{e:rk-1}
\rr^{k}=&\y-\A[S_{k}]\hat{\x}[S_{k}]\nonumber \\
=&\big(\I-\A[S_{k}](\A^\top[S_{k}]\A[S_{k}])^{-1}\A^\top[S_{k}]\big)\y\nonumber \\
\overset{(a)}{=}&\P^{\perp}[S_{k}](\A\x+\v)\nonumber \\
\overset{(b)}{=}&\P^{\perp}[S_{k}](\A[\Omega]\x[\Omega]+\v)\nonumber \\
\overset{(c)}{=}&\P^{\perp}[S_{k}] (\A[S_{k}]\x[S_{k}]+\A[\Omega\setminus S_{k}]\x[\Omega\setminus S_{k}]+\v) \nonumber \\
\overset{(d)}{=}&\P^{\perp}[S_{k}]\A[\Omega\setminus S_{k}]\x[\Omega\setminus S_{k}]+\P^{\perp}[S_{k}]\v,
\end{align}
where (a) is from the definition of $\P^{\perp}[S_{k}]$,
(b) is from the definition of $\Omega$,
(c) follows from the induction assumption that $S_{k}\subseteq \Omega$
and (d) is from $\P^{\perp}[S_{k}]\A[S_{k}]=\0$.

Thus, by~\eqref{e:rk-1}, we obtain
\begin{align}
\label{e:left}
\|\A^\top[\Omega\setminus S_{k}]\rr^{k}]\|_{2,\infty}
\geq \,\;\|\A^\top[\Omega\setminus S_{k}]\P^{\perp}[S_{k}]\A[\Omega\setminus S_{k}]\x[\Omega\setminus S_{k}]\|_{2,\infty}
-\|\A^\top[\Omega\setminus S_{k}]\P^{\perp}[S_{k}]\v\|_{2,\infty},
\end{align}
for the left-hand side of \eqref{e:generalcond}, and for the right-hand side of \eqref{e:generalcond}, we have
\begin{align}
\label{e:right}
\|\A^\top[\Omega^c]\rr^{k}]\|_{2,\infty}
\leq\,\;\|\A^\top[\Omega^c]\P^{\perp}[S_{k}]\A[\Omega\setminus S_{k}]\x[\Omega\setminus S_{k}]\|_{2,\infty}
&+\|\A^\top[\Omega^c]\P^{\perp}[S_{k}]\v\|_{2,\infty}.
\end{align}
Therefore, by~\eqref{e:left} and~\eqref{e:right}, in order to show~\eqref{e:generalcond}, we show
\begin{align}
&\|\A^\top[\Omega\setminus S_{k}]\P^{\perp}[S_{k}]\A[\Omega\setminus S_{k}]\x[\Omega\setminus S_{k}]\|_{2,\infty} -\|\A^\top[\Omega^c]\P^{\perp}[S_{k}]\A[\Omega\setminus S_{k}]\x[\Omega\setminus S_{k}]\|_{2,\infty} \nonumber\\
>& \|\A^\top[\Omega\setminus S_{k}]\P^{\perp}[S_{k}]\v\|_{2,\infty}+\|\A^\top[\Omega^c]\P^{\perp}[S_{k}]\v\|_{2,\infty}
\label{e:generalcond1}.
\end{align}

By the induction assumption, $S_{k}\subseteq \Omega$,
thus
\begin{align}
\label{e:minxk-1}
\|\x[\Omega\setminus S_{k}]\|_2&\geq \sqrt {|\Omega|-k }\min_{i\in\Omega\setminus S_{k}}\|\x[i]\|_2
\geq \sqrt {|\Omega|-k }\min_{i\in\Omega}\|\x[i]\|_2
\geq \sqrt {f(|\Omega|-k )}\min_{i\in\Omega}\|\x[i]\|_2,
\end{align}
where the last inequality follows from the fact that $f(t)\leq t$ for $t>0$.

In the following, we shall give a lower bound on the left-hand side of~\eqref{e:generalcond1} and
an upper bound on the right-hand side of \eqref{e:generalcond1}, respectively.
Since $S_{k}\subseteq \Omega$ and $|S_{k}|=k$,  by Lemma~\ref{l:main} and \eqref{e:l1l2inequalityG}, we have
\begin{align}
\,\;&\|\A^\top[\Omega\setminus S_{k}]\P^{\perp}[S_{k}]\A[\Omega\setminus S_{k}]\x[\Omega\setminus S_{k}]\|_{\infty}-\|\A^\top[\Omega^c]\P^{\perp}[S_{k}]\A[\Omega\setminus S_{k}]\x[\Omega\setminus S_{k}]\|_{\infty} \nonumber\\
\geq&\frac{(1-\sqrt {f(|\Omega|-k )+1 }\delta_{|\Omega|+1})\|\x[\Omega\setminus S_{k}]\|_2}{\sqrt{f(|\Omega|-k )}}
\overset{(a)}{\geq}\frac{(1-\sqrt {f(K)+1 }\delta_{|\Omega|+1})\|\x[\Omega\setminus S_{k}]\|_2}{\sqrt{f(|\Omega|-k )}}\nonumber\\
\overset{(b)}{\geq}&\frac{(1-\sqrt {f(K)+1 }\delta_{K+1})\|\x[\Omega\setminus S_{k}]\|_2}{\sqrt{f(|\Omega|-k )}}
\overset{(c)}{\geq}(1-\sqrt {f(K)+1 }\delta_{K+1})\min_{i\in\Omega}\|\x[i]\|_2,
\label{e:l2cond1left}
\end{align}
where (a) is because $k\geq1$, $\x$ is block $K$-sparse (i.e., $|\Omega|\leq K$) and $f(t)$ is nondecreasing,
(b) follows from Lemma~\ref{l:monot}, and (c) follows from~\eqref{e:deltaG} and~\eqref{e:minxk-1}.

To give an upper bound on the right-hand side of~\eqref{e:generalcond1}, we notice that there exist
$i_0\in\Omega\setminus S_{k}$ and $j_0\in\Omega^c$ such that
\begin{align*}
\|\A^\top[\Omega\setminus S_{k}]\P^{\perp}[S_{k}]\v\|_{2,\infty}&=\|\A^\top[i_0]\P^{\perp}[S_{k}]\v\|_2,\\
\|\A^\top[\Omega^c]\P^{\perp}[S_{k}]\v\|_{2,\infty}&=\|\A^\top[j_0]\P^{\perp}[S_{k}]\v\|_2.
\end{align*}
Therefore, we have
\begin{align}
\label{e:l2cond1right}
\,&\|\A^\top[\Omega\setminus S_{k}]\P^{\perp}[S_{k}]\v\|_{2,\infty}
+\|\A^\top[\Omega^c]\P^{\perp}[S_{k}]\v\|_{2,\infty}\nonumber \\
=&\|\A^\top[i_0]\P^{\perp}[S_{k}]\v\|_2+\|\A^\top[j_0]\P^{\perp}[S_{k}]\v\|_2\nonumber \\
=&\|\A^\top[i_0\cup j_0]\P^{\perp}[S_{k}]\v\|_{2,1}\nonumber \\
\overset{(a)}{\leq}&\sqrt{2}\|\A^\top[i_0\cup j_0]\P^{\perp}[S_{k}]\v\|_2\nonumber \\
\overset{(b)}{\leq} &\sqrt{2(1+\delta_{K+1})}\|\P^{\perp}[S_{k}]\v\|_2\nonumber \\
\overset{(c)}{\leq} &\sqrt{2(1+\delta_{K+1})}\epsilon,
\end{align}
where (a) is from \eqref{e:l2pnorm} and the fact that $\A^\top[i_0\cup j_0]\P^{\perp}[S_{k}]\v$ is a
$2\times1$ block vector, (b) follows from Lemma~\ref{l:AtRIP} and (c) is because
\beq
\label{e:orthcompv}
\|\P^{\perp}[S_{k}]\v\|_2\leq\|\P^{\perp}[S_{k}]\|_2\|\v\|_2\leq\|\v\|_2\leq\epsilon.
\eeq

From~\eqref{e:l2cond1left} and~\eqref{e:l2cond1right}, we can see that ~\eqref{e:generalcond1}
(or equivalently~\eqref{e:generalcond}) is guaranteed by
\begin{align*}
(1-\sqrt {f(K)+1 }\delta_{K+1})\min_{i\in\Omega}\|\x[i]\|_2> \sqrt{2(1+\delta_{K+1})}\epsilon,
\end{align*}
i.e., by \eqref{e:deltaG},
\beqnn
\min_{i\in\Omega}\|\x[i]\|_2> \frac{\sqrt{2(1+\delta_{K+1})}\epsilon}{1-\sqrt {f(K)+1 }\delta_{K+1}}.
\eeqnn
Thus, if~\eqref{e:suffgenenoise} holds, then the BOMP algorithm selects a correct index in the $k$th iteration.

Next we need to show that the BOMP algorithm performs exactly $|\Omega|$ iterations,
which is equivalent to show that $ \|\rr^k\|_2>\epsilon$ for $1\leq k<|\Omega|$ and
$ \|\rr^{|\Omega|}\|_2\leq\epsilon$.
Since the BOMP algorithm selects a correct index in each iteration under~\eqref{e:suffgenenoise},
by~\eqref{e:rk-1}, for $1\leq k<|\Omega|$, we have
\begin{align}
\|\rr^k\|_2&= \|\P^{\perp}[S_{k}]\A[\Omega\setminus S_{k}]\x[\Omega\setminus S_{k}]+\P^{\perp}[S_{k}]\v\|_2\nonumber\\
&\geq \|\P^{\perp}[S_{k}]\A[\Omega\setminus S_{k}]\x[\Omega\setminus S_{k}]\|_2-\|\P^{\perp}[S_{k}]\v\|_2\nonumber\\
&\overset{(a)}{\geq}\|\P^{\perp}[S_{k}]\A[\Omega\setminus S_{k}]\x[\Omega\setminus S_{k}]\|_2-\epsilon
\nonumber\\
&\overset{(b)}{\geq}\sqrt{1-\delta_{|\Omega|}}\|\x[\Omega\setminus S_k]\|_2-\epsilon\nonumber\\
&\overset{(c)}{\geq}\sqrt{1-\delta_{K+1}}\min_{i\in\Omega}\|\x[i]\|_2-\epsilon,
\label{e:rklbd}
\end{align}
where (a), (b) and (c) are respectively from~\eqref{e:orthcompv}, Lemma~\ref{l:orthogonalcomp},
and Lemma~\ref{l:monot}. Therefore, if
\beq
\label{e:earlycondl2}
\min_{i\in\Omega}\|\x[i]\|_2> \frac{2\epsilon}{\sqrt {1-\delta_{K+1}}},
\eeq
then $\|\rr^k\|_2>\epsilon$ for $1\leq k< \Omega$.

By a simple calculation, we can show that
\beq
\label{e:condl22}
\frac{2\epsilon}{1-\sqrt {f(K)+1 }\delta_{K+1}}\geq \frac{2\epsilon}{\sqrt {1-\delta_{K+1}}}.
\eeq
Indeed, we have
\beqnn
1-\sqrt {f(K)+1 }\delta_{K+1}\leq 1-\delta_{K+1}\leq\sqrt {1-\delta_{K+1}}.
\eeqnn
Thus,~\eqref{e:condl22} follows.
Therefore, by~\eqref{e:earlycondl2} and~\eqref{e:condl22}, if~\eqref{e:suffgenenoise} holds, then
$\|\rr^k\|_2>\epsilon$ for $1\leq k< \Omega$, i.e., the BOMP algorithm does not terminate
before the $|\Omega|$-th iteration.

Similarly, by~\eqref{e:rk-1}, we have
\begin{align*}
\|\rr^{|\Omega|}\|_2&= \|\P^{\perp}_{S_{|\Omega|}}\A_{\Omega\setminus S_{|\Omega|}}
\x_{\Omega\setminus S_{|\Omega|}}+\P^{\perp}[S_{|\Omega|}]\v\|_2
\overset{(a)}{=}\|\P^{\perp}_{S_{|\Omega|}}\v\|_2
\overset{(b)}{\leq}\epsilon,
\end{align*}
where (a) is because $S_{|\Omega|}=|\Omega|$ and (b) follows from~\eqref{e:orthcompv}.
So, by the stopping condition, the BOMP algorithm terminates after
performing the  $|\Omega|$-th iteration in total.
That is, the BOMP algorithm performs $|\Omega|$ iterations.

Next let us prove the second part, i.e. \eqref{e:xerr} of Theorem~\ref{t:SDG}.
Since the support of the block $K$-sparse signal $\x$ can be exactly recovered by the BOMP algorithm, we get
\begin{align*}
\hat{\x}&=(\A^\top[\Omega]\A[\Omega])^{-1}\A^\top[\Omega]\y
=(\A^\top[\Omega]\A[\Omega])^{-1}\A^\top[\Omega](\A\x+\v)=\x+\P[\Omega]\v.
\end{align*}
Thus, we have
\[
\|\x-\hat{\x}\|_2=\|\P[\Omega]\v\|_2\leq\|\v\|_2\leq \epsilon.
\]
This completes the proof. $\Box$

\subsection{Proof of Theorem~\ref{t:NR}}
\label{ss:pfTNR}
{\em Proof.}
For any given positive integers $d, K\geq 1$, let matrix function
\begin{equation*}
\B(d)=\begin{bmatrix}
\frac{K}{K+1}\I_{dK}&  \frac{\E_{(dK)\times d}}{K+1} \\ \frac{\E^\top_{(dK)\times d}}{K+1} & \frac{K+2}{K+1}\I_{d}
\end{bmatrix},
\end{equation*}
where
\beq
\label{e:E}
\E_{(dK)\times d}=(\I_d,\ldots, \I_d)^\top \in \mathbb{R}^{(dK)\times d}
\eeq
with $\I_{d}$ being the $d\times d$ identity matrix.

Let
\beq
\label{e:sCd}
s=\delta-\frac{1}{\sqrt{K+1}},\quad \C(d)=\B(d)-s\I_{d(K+1)}
\eeq
and
\begin{equation}
\label{keyvector}
\x(d)=(\1_d^\top,\ldots,\1_d^\top,\0_d^\top)\in \mathbb{R}^{d(K+1)},
\end{equation}
where $\1_d\in \mathbb{R}^d$ with all of its entries being 1. Then $\x(d)$ is a block $K$-sparse signal.

In the following, we plan to show that $\C(d)$ is symmetric positive definite.
Then  there exists an upper triangular matrix $\A(d)$ such that $\A^\top(d)\A(d)=\C(d)$.
We will continue to show that $\A(d)$ satisfies the block-RIP of order $K+1$ with $\delta_{K+1}=\delta$,
and the BOMP algorithm may fail to recover the block $K$-sparse signal $\x(d)$
from $\y(d)=\A(d)\x(d)$ in $K$ iterations.

By some tedious calculations, it is not hard to show that the eigenvalues
$\{\lambda_i\}_{i=1}^{K+1}$ of $\B(1)$ are
\begin{align*}
\lambda_1=\ldots=\lambda_{K-1}=\frac{K}{K+1},
\lambda_{K}=1-\frac{1}{\sqrt{K+1}}, \quad \lambda_{K+1}=1+\frac{1}{\sqrt{K+1}}.
\end{align*}

To show $\C(d)$ is positive definite, we claim that for each $\x\in \mathbb{R}^{(d+1)K}$,
\begin{align}
\label{e:eig}
\left(1-\frac{1}{\sqrt{K+1}}\right)\|\x\|_2^2\leq \x^\top\B(d)\x\leq\left(1+\frac{1}{\sqrt{K+1}}\right)\|\x\|_2^2.
\end{align}
Let $\u, \v\in \mathbb{R}^{K+1}$ with $u_i=v_i=\|\x[i]\|_2$ for $1\leq i\leq K$ and
$u_{K+1}=-v_{K+1}=\|\x[K+1]\|_2$.
Then
\begin{align*}
\x^\top\B(d)\x
\overset{(a)}{=}&\frac{K}{K+1}\sum_{i=1}^K\x^\top[i]\x[i]+\frac{K+2}{K+1}\x^\top[K+1]\x[K+1]
+\frac{2}{K+1}\sum_{i=1}^K(\x^\top[i]\x[K+1])\\
\overset{(b)}{\leq}&\frac{K}{K+1}\sum_{i=1}^K\|\x^\top[i]\|^2+\frac{K+2}{K+1}\|\x^\top[K+1]\|_2^2
+\frac{2}{K+1}\sum_{i=1}^K(\|\x^\top[i]\|_2\|\x[K+1]\|_2)\\
=&\frac{K}{K+1}\sum_{i=1}^Ku_i^2+\frac{K+2}{K+1}u_{K+1}^2+\frac{2}{K+1}\sum_{i=1}^K(u_iu_{K+1})\\
=&\u^\top\B(1)\u\leq\left(1+\frac{1}{\sqrt{K+1}}\right)\|\u\|_2^2=\left(1+\frac{1}{\sqrt{K+1}}\right)\|\x\|_2^2,
\end{align*}
where (a) and (b) respectively follow from \eqref{e:E} and the Cauchy-Schwartz  inequality.
Similarly,
\begin{align*}
\x^\top\B(d)\x
\geq &\frac{K}{K+1}\sum_{i=1}^K\|\x^\top[i]\|^2+\frac{K+2}{K+1}\|\x^\top[K+1]\|_2^2
-\frac{2}{K+1}\sum_{i=1}^K(\|\x^\top[i]\|_2\|\x[K+1]\|_2)\\
=&\frac{K}{K+1}\sum_{i=1}^Kv_i^2+\frac{K+2}{K+1}v_{K+1}^2+\frac{2}{K+1}\sum_{i=1}^K(v_iv_{K+1})\\
=&\v^\top\B(1)\v\geq\left(1-\frac{1}{\sqrt{K+1}}\right)\|\v\|_2^2 =\left(1-\frac{1}{\sqrt{K+1}}\right)\|\x\|_2^2.
\end{align*}
Thus, \eqref{e:eig} follows. By \eqref{e:sCd} and \eqref{e:eig}, we have
\[
(1-\delta)\|\x\|_2^2\leq \x^\top\C(d)\x\leq\left[1+\left(\frac{2}{\sqrt{K+1}}-\delta\right)\right]\|\x\|_2^2.
\]
By \eqref{e:deltaNR}, we can see that $\C(d)$ is a symmetric positive definite matrix.
Therefore, there exists an upper triangular matrix $\A(d)$ such that $\A^\top(d)\A(d)=\C(d)$.
Furthermore, $\A(d)$ satisfies the block-RIP of order $K+1$ with $\delta_{K+1}\leq \delta$.

Next we need to show that $\A(d)$ satisfies the block-RIP of order $K+1$ with
$\delta_{K+1}\geq \delta$, leading to $\delta_{K+1}=\delta$.
By \eqref{e:eig}, it suffices to show there exists a
$\x\in \mathbb{R}^{(d+1)K}$ such that
\[
\x^\top\B(d)\x=\left(1-\frac{1}{\sqrt{K+1}}\right)\|\x\|_2^2.
\]
Indeed,
in this case, we will have
\begin{align*}
\|\A(d)\x\|_2^2&= \x^\top \C(d)\x= \x^\top (\B(d)-s\I(d)) \x
= (1-\frac{1}{\sqrt{K+1}}-s)\|\x\|_2^2=(1-\delta)\|\x\|^2_2.
\end{align*}

Let $\bar{\u}\in \mathbb{R}^{K+1}$ be the eigenvector of $\B(1)$ corresponding to
the eigenvalue $1-\frac{1}{\sqrt{K+1}}$.
Then
\[
\bar{\u}^\top \B(1)\bar{\u}=\left(1-\frac{1}{\sqrt{K+1}}\right)\|\bar{\u}\|_2^2.
\]
Let $\x\in \mathbb{R}^{d(K+1)}$ satisfying $\x[i]=\bar{u}_i\e_1$
(here $\e_1\in \mathbb{R}^d$ is the first column of the $d\times d$ identity matrix) for $1\leq i\leq K+1$. Then
\[
\x^\top\B(d)\x=\left(1-\frac{1}{\sqrt{K+1}}\right)\|\bar{\u}\|_2^2=\left(1-\frac{1}{\sqrt{K+1}}\right)\|\x\|_2^2.
\]
Thus, $\A(d)$ satisfies the block-RIP of order $K+1$ with $\delta_{K+1}= \delta$.

Finally, we show that BOMP may fail to recover the block $K$-sparse signal $\x(d)$ given in (\ref{keyvector})
from $\y(d)=\A(d)\x(d)$ in $K$ iterations. By the fact that $\A^\top(d)\A(d)=\C(d)$ and
\eqref{keyvector},  we have
\begin{align*}
\max_{j\in \Omega^c}\|\left(\A(d)[j]\right)^\top\y\|_2=&
\|\left(\A(d)[K+1]\right)^\top\y\|_2=\|\left(\A(d)[K+1]\right)^\top\A(d)\x(d)\|_2\\
=&\|(\0_{dK\times d},\I_d)^\top\C(d)\x(d)\|_2=\frac{K\sqrt{d}}{K+1}.
\end{align*}
Similarly, for $1\leq i\leq K$, we have
\begin{align*}
\|\left(\A(d)[i]\right)^\top\y\|_2=&\|\left(\A(d)[i]\right)^\top\A(d)\x(d)\|_2
=\|(\0_{(i-1)K\times d},\I_d,\0_{(K+1-i)K\times d})^\top \A^\top(d) \A(d)\x(d)\|_2\\
=&(\frac{K}{K+1}-s)\sqrt{d}.
\end{align*}
Since $s\geq0$, it follows that
\[
\max_{i\in \Omega}\|\left(\A(d)[i]\right)^\top\y\|_2\le \max_{j\in \Omega^c}\|\left(\A(d)[j]\right)^\top\y\|_2.
\]
Therefore, the BOMP algorithm may fail in the first iteration.
As a result, the BOMP algorithm may fail to recover the block $K$-sparse signal $\x(d)$ in $K$ iterations.
\ \ $\Box$

\subsection{Proof of Lemma~\ref{l:gi}}
\label{ss:pfLg}
By \eqref{e:gi}, we have
\[
g_i(t)=\frac{t+1}{t-1}\big(1-\frac{2}{t^i+1}\big).
\]
Since $t>1$, it is easy to see that \eqref{e:giorder} holds.

In the following, let us show that $g_i(t)$ is strictly decreasing with $t$ for $i\geq2$.
By \eqref{e:gi} and a simple calculation,  we have
\[
g_i(t)=1+2\phi(t) \hbox{ with }
\phi(t)=\frac{(t^i-t)}{t^{i+1}-t^i+t-1}.
\]
Thus, we only need to show $\phi(t)$ is strictly decreasing with $t$ for $i\geq2$.
By direct calculation, we have
\[
\phi'(t)=\frac{\overline{\phi}(t)}{(t^{i+1}-t^i+t-1)^2},
\]
where
\[
\overline{\phi}(t)=(1+it^{i+1})-(t^{2i}+it^{i-1}).
\]
Thus, we only need to show $\overline{\phi}(t)<0$ for $t>1$ and $i\geq2$.
An elementary calculation yields
\begin{align*}
\overline{\phi}'(t)&=i(i+1)t^{i}-(2it^{2i-1}+i(i-1)t^{i-2})
=t^{i-2}i\big[(i+1)t^{2}-(2t^{i+1}+i-1)\big]<0
\end{align*}
for $t>1$. Thus, $\overline{\phi}(t)<\overline{\phi}(1)=0$.
Therefore, $g_i(t)$ is decreasing with $t$ for $i\geq2$. By some simple calculations, one can show that
\[
\lim_{t\rightarrow 1}g_i(t)=i, \quad \lim_{t\rightarrow +\infty}g_i(t)=1.
\]
Thus, $1< g_i(t)< i$ for $i\geq2$. \ \ $\Box$

\subsection{Proof of Theorem \ref{t:SDSRN}}
\label{SS:pfTSDSRN}
Before proving Theorem \ref{t:SDSRN}, we need to introduce the following lemma
which is a variant of \cite[Lemma 1]{DinCG13}.
\begin{lemma}
\label{l:gbar}
Let $\alpha>1$ and  $\beta\geq \gamma\geq1$ be given constants and let
\beq
\label{e:gbari}
\overline{g}_i(t_1,\ldots,t_i, \beta, \gamma)
=\frac{(\sum_{j=1}^{i-1} t_j+\beta t_i)^2}{\sum_{j=1}^{i-1} t_j^2+\gamma t_i^2},
\quad 1\leq i\leq K,
\eeq
where
\beq
\label{e:t}
t_1\geq \alpha t_2\geq\ldots \geq \alpha^{K-1} t_K>0, \quad \sum_{j=1}^{0}\cdot =0.
\eeq
Then, for $1\leq i\leq K$, $\overline{g}_i$ is increasing with $t_i$, and
\beq
\label{e:ggbar}
\overline{g}_i(t_1,\ldots,t_i, 1,1)\leq g_i(\alpha),
\eeq
where function $g_i$ is defined in \eqref{e:gi}.
\end{lemma}

{\em Proof}. We first show that $\overline{g}_i$ is increasing with $t_i$. By \eqref{e:gbari}, we have
\begin{align*}
\frac{\partial \overline{g}_i}{\partial t_i}=&2(\sum_{j=1}^{i-1} t_j+\beta t_i)
\frac{\beta(\sum_{j=1}^{i-1} t_j^2+\gamma t_i^2)-(\sum_{j=1}^{i-1} t_j+\beta t_i)\gamma t_i}{(\sum_{j=1}^{i-1} t_j^2+\gamma t_i^2)^2}.
\end{align*}
By \eqref{e:t}, and the facts that $\alpha>1$ and $\beta\geq \gamma>1$,
it is easy to see that $\frac{\partial \overline{g}_i}{\partial t_i}>0$,
so $\overline{g}_i$ is increasing with $t_i$.

By \eqref{e:giorder}, \eqref{e:ggbar} holds with $i=1$.
For $2\le i\le K$, we have
\begin{align*}
\overline{g}_i(t_1,\ldots,t_i, 1, 1)
\leq & \overline{g}_i(t_1,\ldots,t_{i-1}, \alpha^{-1}t_{i-1},1, 1)
=\overline{g}_{i-1}(t_1,\ldots,t_{i-1},(1+\alpha^{-1}), (1+\alpha^{-2}))\\
\leq&\ldots\leq \overline{g}_1(t_1,(\sum_{j=1}^{i}\alpha^{1-j}), (\sum_{j=1}^{i}\alpha^{2-2j}))=g_i(\alpha).
\end{align*}
Thus, \eqref{e:ggbar} hold for $1\le i\le K$.\ \ $\Box$

{\em Proof of Theorem \ref{t:SDSRN}.}
If $s^{-2}-1\geq K$, then $s\leq\frac{1}{\sqrt{1+K}}$. By \eqref{e:s}, $\delta_{K+1}<\frac{1}{\sqrt{1+K}}$.
Thus, by Corollary \ref{c:SRN}, Theorem \ref{t:SDSRN} holds.

Next we consider the case that $s^{-2}-1< K$. By \eqref{e:ginverse}, we have
\[
g_{K}(g_{K}^{-1}(s^{-2}-1))=s^{-2}-1.
\]
Let $f(t)=\min\{s^{-2}-1,t\}$. By  Theorem \ref{t:SDG},
it suffices to show \eqref{e:l1l2inequalityG} holds under \eqref{e:s} and \eqref{e:alphaCSD}.

By Definition~\ref{d:blockstrong}, if $\x$ is a block $\alpha$-strongly-decaying $K$-sparse signal,
so is $\x[\Omega \setminus S]$ for  any subset $S$  of $\Omega$.
By \eqref{e:gbari},
\begin{align*}
\frac{\|\x[\Omega \setminus S]\|_{2,1}^2}{\|\x[\Omega \setminus S]\|_{2}^2}
=&\overline{g}_{|\Omega\setminus S|}\big(\|(\x[\Omega \setminus S])[1]\big\|_2,\ldots,\big\|(\x[\Omega \setminus S])[|\Omega\setminus S|]\big\|_2, 1, 1\big)\\
\overset{(a)}{\leq}&g_{|\Omega\setminus S|}(\alpha)
\overset{(b)}{<}g_{|\Omega\setminus S|}(g_{K}^{-1}(s^{-2}-1))
\overset{(c)}{<} g_{K}(g_{K}^{-1}(s^{-2}-1))=s^{-2}-1,
\end{align*}
where (a) follows from \eqref{e:ggbar} and Definition \ref{d:blockstrong},
(b) is from \eqref{e:alphaCSD} and Lemma \ref{l:gi}, and (c) follows from \eqref{e:giorder}.

By the Cauchy-Schwarz inequality, obviously,
\[
\|\x[\Omega \setminus S]\|_{2,1}^2\leq |\Omega \setminus S| \|\x[\Omega \setminus S]\|_{2}^2.
\]
Thus, the above two inequalities imply that \eqref{e:l1l2inequalityG} holds with $f(t)=\min\{s^{-2}-1,t\}$.
Therefore, Theorem \ref{t:SDSRN} follows from Theorem \ref{t:SDG}.
\ \ $\Box$

\subsection{Proof of Theorem \ref{t:SDNR}}
\label{SS:pfTSDNR}
{\em Proof.}  For any given positive integers $d, K\geq 1$, $\delta$ satisfies \eqref{e:deltaSDNR}, and $\alpha>1$, let
\begin{equation*}
\B(d)=\begin{bmatrix}
\frac{1+\sqrt{2}-2\delta}{2}\I_{d}&  \frac{1}{2}\I_{d}&\0_{d\times d(K-1)}\\
\frac{1}{2}\I_{d}&  \frac{3+\sqrt{2}-2\delta}{2}\I_{d}&\0_{d\times d(K-1)}\\
\0_{d(K-1)\times d}& \0_{d(K-1)\times d}&\frac{1+\sqrt{2}-2\delta}{2}\I_{d(K-1)}
\end{bmatrix}
\end{equation*}
and
\begin{align*}
\x(d)=(a_1\1_d;\0_d;a_2\1_d;\ldots; a_K\1_d)^\top \in \mathbb{R}^{d(K+1)},
\end{align*}
where all the entries of $\1_d\in \mathbb{R}^d$ are 1, and
\[
a_1\geq \alpha a_2\geq \ldots\geq \alpha^{K-1} a_K.
\]
Thus, $\x(d)$ is a block $\alpha$-strongly-decaying $K$-sparse signal.

Following the proof of Theorem \ref{t:NR}, one can easily show that $\B(d)$ is symmetric positive definite.
Thus, there exists an upper triangular matrix $\A(d)$ such that $\A^\top(d)\A(d)=\B(d)$.
Moreover, one can similarly show  that $\A(d)$ satisfies the block-RIP of order $K+1$ with $\delta_{K+1}=\delta$.

In the following, we show that BOMP may fail to recover the block $K$-sparse signal
$\x(d)$ from $\y(d)=\A(d)\x(d)$ in $K$ iterations.
By the fact that $\A^\top(d)\A(d)=\B(d)$, we have
\begin{align*}
\max_{j\in \Omega^c}\|\A^\top(d)[j]\y\|_2=\|\A^\top(d)[2]\y\|_2
=\|(\0_{d\times d},\I_d,\0_{d(K-1)\times d})\A^\top(d)\A(d)\x(d)\|_2
=\frac{1}{2}a_1\sqrt{d}.
\end{align*}
Similarly, for $i\in \Omega$, we have
\begin{align*}
\|\A^\top(d)[i]\y\|_2&=\|\A^\top(d)[i]\A(d)\x(d)\|_2
=\frac{1+\sqrt{2}-2\delta}{2}a_i\sqrt{d}.
\end{align*}
Thus, by \eqref{e:deltaSDNR}, it holds that
\begin{align*}
\max_{i\in \Omega}\|\A^\top(d)[i]\y\|_2&=\|\A^\top(d)[1]\y\|_2
\leq \max_{j\in \Omega^c}\|\A^\top(d)[j]\y\|_2.
\end{align*}
Therefore, BOMP may fail in the first iteration.
As a result, the BOMP algorithm may fail to recover the block $\alpha$-strongly-decaying $K$-sparse signal
$\x(d)$ in $K$ iterations.
\ \ $\Box$

\subsection{Proof of Theorem \ref{t:SDSRNP}}
\label{SS:pfTSDSRNP}
Before proving Theorem \ref{t:SDSRNP}, we need to introduce the following lemma.
\begin{lemma}
\label{l:hbar}
Let $\beta\geq1$ be a given constant and let
\beq
\label{e:hbari}
\overline{h}_i(t_1,\ldots,t_i, \beta)=\frac{(\sum_{j=1}^{i-1} t_j
+\beta t_i)^2}{\sum_{j=1}^{i-1} t_j^2+\beta t_i^2}, \quad 1\leq i\leq K,
\eeq
where
\beq
\label{e:talpha}
t_1\geq \alpha_1 t_2\geq\ldots \geq \alpha_{K-1} t_K>0, \quad \sum_{j=1}^{0}\cdot =0
\eeq
for given $\alpha_1,\ldots, \alpha_{K-1}\geq1$.
Then, $\overline{h}_i$ is increasing with $t_i$. Moreover, for $2\leq i\leq K$,
\beq
\label{e:hhbat}
\overline{h}_i(t_1,\ldots,t_i, 1)=\hat{h}_{i-1}(t_1/t_2)\leq \hat{h}_{i-1}(\alpha_1),
\eeq
where $\hat{h}_{i-1}$ is defined in \eqref{e:hat}.
\end{lemma}
{\em Proof}. By \eqref{e:gbari} and \eqref{e:hbari}, we have
\[
\overline{h}_i(t_1,\ldots,t_i, \beta)=\overline{g}_i(t_1,\ldots,t_i, \beta, \beta).
\]
By Lemma \ref{l:gbar}, $\overline{h}_i$ is increasing with $t_i$. Next we prove \eqref{e:hhbat}.

Since  $\overline{h}_i$ is increasing with $t_i$, by \eqref{e:talpha} and the fact that $\alpha_i\geq1$ for $1\leq i\leq K-1$, we have
\begin{align*}
\overline{h}_i(t_1,\ldots,t_i, 1)
\leq& \overline{h}_i(t_1,\ldots,t_{i-1}, t_{i-1},1)=\overline{h}_{i-1}(t_1,\ldots,t_{i-1},2)\\
\leq&\ldots\leq \overline{h}_2(t_1,t_2, i-1)=\hat{h}_{i-1}(t_1/t_2)\leq \hat{h}_{i-1}(\alpha_1),
\end{align*}
where the last inequality follows from Lemma \ref{l:hat}.\ \ $\Box$

{\em Proof of Theorem \ref{t:SDSRNP}.}
We prove it by induction.  We first show that the BOMP algorithm chooses an index in $\Omega$
in the first iteration.

If $s^{-2}-1\geq K$, then $s\leq\frac{1}{\sqrt{1+K}}$. By \eqref{e:spartial},
$\delta_{K+1}<\frac{1}{\sqrt{1+K}}$.
Thus, by Corollary \ref{c:SRN}, the BOMP algorithm chooses an index in $\Omega$ in the first iteration.

If $s^{-2}-1< K$, we let $\alpha_1=\|\x[1]\|_2/\|\x[2]\|_2$ and $f(t)=\min\{s^{-2}-1,t\}$.
Thus, by Theorem \ref{t:SDSRN}, to show the BOMP algorithm chooses an index in $\Omega$
in the first iteration, it suffices to show \eqref{e:l1l2inequalityG}
holds with $S=\emptyset$ under \eqref{e:alphaCSDP}, which is equivalent to show
\[
\frac{\|\x\|_{2,1}^2}{\|\x\|_{2}^2}\leq s^{-2}-1.
\]
By Lemma~\ref{l:hbar} with $\alpha_2=\ldots=\alpha_{K-1}=1$, Lemma~\ref{l:hat}, and
\eqref {e:hat}-\eqref{e:spartial}, we have
\begin{align*}
\frac{\|\x\|_{2,1}^2}{\|\x\|_{2}^2}
=\overline{h}_{K}\big(\|\x[1]\|_2,\ldots,\|\x[K]\|_2, 1\big)
\leq \hat{h}_{K-1}(\alpha_1)\leq \hat{h}_{K-1}(h_{K-1}(s^{-2}-1))= s^{-2}-1.
\end{align*}

Let us assume that the BOMP algorithm chooses an index in $\Omega$ in the first $k$ iterations.
Then, we show the BOMP chooses an index in $\Omega$ in the  $(k+1)$-th iteration ($1\leq k\leq i-1$).

Suppose $S$ is the set of all the indexes chosen by the BOMP algorithm in the first $k$ iterations.
By Theorem \ref{t:SDG}, it suffices to
show \eqref{e:l1l2inequalityG} holds under \eqref{e:alphaCSDP}, which is equivalent to show
\beq
\label{e:partialcond}
\frac{\|\x[\Omega \setminus S]\|_{2,1}^2}{\|\x[\Omega \setminus S]\|_{2}^2}\leq f(|\Omega \setminus S|)=\min\{s^{-2}-1,|\Omega \setminus S|\}.
\eeq
By the Cauchy-Schwartz inequality and \eqref{e:l2pnorm}, we have
\[
\|\x[\Omega \setminus S]\|_{2,1}^2\leq (|\Omega \setminus S|)\|\x[\Omega \setminus S]\|_{2}^2.
\]
Thus, if $s^{-2}-1\geq  K-|S|\geq |\Omega \setminus S|$, then \eqref{e:partialcond} holds.
Hence, we can use Theorem \ref{t:SDG}.

It remains to consider the case $s^{-2}-1<  K-|S|$. Let us prove
\[
\frac{\|\x[\Omega \setminus S]\|_{2,1}^2}{\|\x[\Omega \setminus S]\|_{2}^2}\leq s^{-2}-1.
\]
To this end, let
\[
\alpha_1=\min_{1\leq j\leq k+1}\frac{\|\x[j]\|_2}{\|\x[j+1]\|_2}.
\]
Then, we have
\beq
\label{e:partialcond2}
\|(\x[\Omega \setminus S])[1]\|_2\geq \alpha_1\|(\x[\Omega \setminus S])[2]\|_2.
\eeq
And by \eqref{e:hi}-\eqref{e:spartial}, we have
\[
\alpha_1\geq h_{K-1}(s^{-2}-1)\geq h_{K-(k+1)}(s^{-2}-1).
\]

By Lemma~\ref{l:hbar} with $\alpha_2=\ldots=\alpha_{K-1}=1$, Lemma~\ref{l:hat}, and
\eqref {e:hat}-\eqref{e:spartial}, we have
\begin{align*}
\frac{\|\x[\Omega \setminus S]\|_{2,1}^2}{\|\x[\Omega \setminus S]\|_{2}^2}
=&\overline{h}_{K-k}\big(\|(\x[\Omega \setminus S])[1]\|_2,\ldots,\|(\x[\Omega \setminus S])[K-k]\|_2, 1\big)\\
\leq& \hat{h}_{K-(k+1)}(\alpha_1)\leq \hat{h}_{K-(k+1)}(h_{K-(k+1)}(s^{-2}-1))=s^{-2}-1.
\end{align*}
Thus, with \eqref{e:partialcond2}, we have \eqref{e:partialcond} and hence, we can apply Theorem~\ref{t:SDG}.
These complete the proof. \ \ $\Box$

\section{Conclusions and Future Work}
\label{s:con}
In this paper, we studied block-RIC based sufficient and necessary conditions for
exact  and stable recovery of block $K$-sparse signals $\x$ from the measurements $\y=\A\x+\v$ by the BOMP algorithm in $K$ iterations,
where $\v$ is an $\ell_2$-bounded noise vector.
First, we showed that $\delta_{K+1}<1/\sqrt{K+1}$ is a sharp sufficient condition.
Second, under $\delta_{K+1}<\sqrt{2}/2$, we presented a class of
block $K$-sparse signals which can be recovered exactly or stably by using the BOMP algorithm in $K$ iterations.
That is, we showed that under some conditions on $\alpha$,  the BOMP algorithm can exactly or stably recover
any block $\alpha$-strongly-decaying $K$-sparse signal in $K$ iterations
when $\delta_{K+1}<\sqrt{2}/2$.
Moreover, we proved that the condition on $\alpha$ is sharp.
Finally, we showed that $\delta_{K+1}<\sqrt{2}/2$ is a  sufficient condition for partial recovery of
block $K$-sparse signals $\x$ by the BOMP algorithm
under certain condition on the signals $\x$.
The condition on the blocks of block $K$-sparse signals $\x$ is also shown to  be sharp for partial recovery.

In the following, we point out some research problems that will be investigated in the future
\begin{enumerate}

\item It is possible to use more than $K$ iterations of OMP  to
recover a sparse signal as discussed in \cite{Zha11, Liv12, LivT14,
CohDD17, WanS16} or choose more than
one indexes per iteration (see \cite{HuaZ11,LiuT12, WanKS12b}).
It can be expected that less restrictive  sufficient conditions can be obtained,
and we will investigate it in the future.

\item We are interested in extending our current work to this setting of study.
For example,  besides $\alpha$-strongly-decaying sparse signals, what other signals can be recovered exactly
or stably by using the BOMP in $K$ iterations?

\item We assume that all the blocks have identical size of $d$ in this paper. As a
future work, we will extend our work to the case with non-uniform
block sizes. Moreover, we are interested in investigating
sufficient conditions of recovering
$\alpha$-strongly-decaying sparse signals by solving the $\ell_1$
-minimization problem (see \eqref{e:l1}).

\item An extension of the BOMP algorithm called as model-aware orthogonal matching pursuit (MA-OMP)
has recently been proposed in \cite{WeiWU15} to recover the coherent MIMO radar imaging.
Exact recovery of block $K$-sparse signals with structure by  MA-OMP has been investigated
in \cite{WisWU16}. Whether the techniques developed in this paper can be used to develop a
sharp condition for MA-OMP will be studied in the future.
\end{enumerate}

\section{Acknowledgment}
This work was partially supported by the National Science Foundation of China under Grant 61661146003, 
the Joint fund of the Ministry of Education under Grant 6141A020223, Simons collaboration grant 
(No. 280646) and the National Science Foundation under the grant (No. DMS 1521537),
and Major Frontier Project of Sichuan Province under Grant 2015JY0282.

Part of this work was carried out while Jinming Wen and Ming-Jun Lai were visiting
Zhengchun Zhou and Xiaohu Tang at Southwest Jiaotong University.
The hospitality received in this period is gratefully acknowledged.

\bibliographystyle{SIAM}
\bibliography{ref-RIP2}
\end{document}